\documentclass[lettersize,journal]{IEEEtran}
\usepackage{times}
\usepackage{graphicx}
\usepackage{amsmath, amsfonts}
\usepackage{amssymb}
\usepackage{amsthm}
\newtheorem{proposition}{Proposition}
\usepackage{algorithm}
\usepackage{algpseudocode}
\usepackage{multirow}
\usepackage{bm}
\usepackage{float}
\usepackage{lettrine}
\usepackage{stfloats}
\usepackage{cite}
\usepackage{amsmath}
\usepackage{diagbox}
\usepackage{makecell}
\usepackage{subfigure}
\usepackage{caption}
\usepackage{subfig} 
\usepackage{xcolor}
\usepackage{url}
\usepackage{nomencl}
\usepackage{colortbl}
\makenomenclature

\usepackage{enumitem}
\usepackage{booktabs}
\usepackage{tabularx}
\usepackage{xspace}

\begin{document}

\title{MARS: Modality-Aligned Retrieval for Sequence Augmented CTR Prediction}

\author{Yutian Xiao, Shukuan Wang, Zhao Zhang\IEEEauthorrefmark{1}, Meng Yuan, Wei Chen, Hao Geng, Chenghao Zhang, Yanze Zhang, Shanqi Liu, Chao Feng, Xiang Li, Lantao Hu, Han Li, Fuzhen Zhuang\IEEEauthorrefmark{1}, \IEEEmembership{Senior Member,~IEEE}
\IEEEcompsocitemizethanks{
\IEEEcompsocthanksitem Yutian Xiao, Meng Yuan, Wei Chen, Fuzhen Zhuang are with the School of Artificial Intelligence in Beihang University, Beijing, China. (e-mail: by2442221@buaa.edu.cn, yuanmeng97@buaa.edu.cn, cwei\_01@163.com, zhuangfuzhen@buaa.edu.cn)
\IEEEcompsocthanksitem Shukuan Wang, Chenghao Zhang, Yanze Zhang, Shanqi Liu, Chao Feng, Xiang Li, Lantao Hu, Han Li are with the Kuaishou Technology Co., Ltd., Beijing, China. (e-mail: wangshukuan@kuaishou.com, zhangchenghao03@kuaishou.com, zhangyanze@kuaishou.com, liushanqi@kuaishou.com, fengchao08@kuaishou.com, lixiang44@kuaishou.com, hulantao@kuaishou.com, lihan08@kuaishou.com)
\IEEEcompsocthanksitem Zhao Zhang, Hao Geng are with the School of Computer Science and Engineering in Beihang University, Beijing, China. (e-mail: zhao\_zhang@buaa.edu.cn, genghao@buaa.edu.cn.)
}

\thanks{\IEEEauthorrefmark{1} indicates the corresponding authors. \IEEEauthorrefmark{2} indicates the equal contribution.}}

\markboth{IEEE Transactions on Knowledge and Data Engineering}
{Shell \MakeLowercase{\textit{et al.}}: IEEE Transactions on Knowledge and Data Engineering}

\IEEEcompsoctitleabstractindextext{

\begin{abstract}
Click-through rate (CTR) prediction serves as a cornerstone of recommender systems. Despite the strong performance of current CTR models based on user behavior modeling, they are still severely limited by interaction sparsity, especially in low-active user scenarios. To address this issue, data augmentation of user behavior is a promising research direction. However, existing data augmentation methods heavily rely on collaborative signals while overlooking the rich multimodal features of items, leading to insufficient modeling of low-active users.

To alleviate this problem, we propose a novel framework \textbf{MARS} (\textbf{M}odality-\textbf{A}ligned \textbf{R}etrieval for \textbf{S}equence Augmented CTR Prediction). MARS utilizes a Stein kernel-based approach to align text and image features into a unified and unbiased semantic space to construct multimodal user embeddings. Subsequently, each low-active user's behavior sequence is augmented by retrieving, filtering, and concentrating the most similar behavior sequence of high-active users via multimodal user embeddings.
Validated by extensive offline experiments and online A/B tests, our framework MARS consistently outperforms state-of-the-art baselines and achieves substantial growth on core business
metrics for low-active users on Kuaishou~\footnote{https://www.kuaishou.com/}. Consequently, MARS has been successfully deployed, serving the main traffic for hundreds of millions of users.
\end{abstract}

}
\maketitle
\IEEEdisplaynotcompsoctitleabstractindextext

\IEEEpeerreviewmaketitle


\section{Introduction}
CTR prediction models have become a fundamental component of modern recommender systems \cite{wang2015collaborative,covington2016deep,qin2023learning,song2019autoint} and online advertising \cite{yang2022click,chen2016deep,gao2023rec4ad,wu2022fedctr}. These models have demonstrated substantial potential in tasks such as ad ranking, personalized recommendation, and user behavior prediction. Early shallow models, such as Logistic Regression (LR) \cite{mcmahan2013ad} and Factorization Machines (FM) \cite{rendle2010factorization}, gained widespread adoption due to their efficiency and interpretability. However, with the advancement of deep learning, deep CTR models such as Wide \& Deep \cite{cheng2016wide}, DIN \cite{zhou2018deep}, and DIEN \cite{zhou2019deep} have continuously pushed the performance boundaries. By leveraging deep neural networks, these models effectively capture complex user behavior patterns and yield remarkable results.

\begin{figure}[t]
  \centering
  \setlength{\fboxrule}{0.pt}
  \setlength{\fboxsep}{0.pt}
  \begin{minipage}[t]{0.49\linewidth}
    \centering
    \fbox{\includegraphics[width=\linewidth]{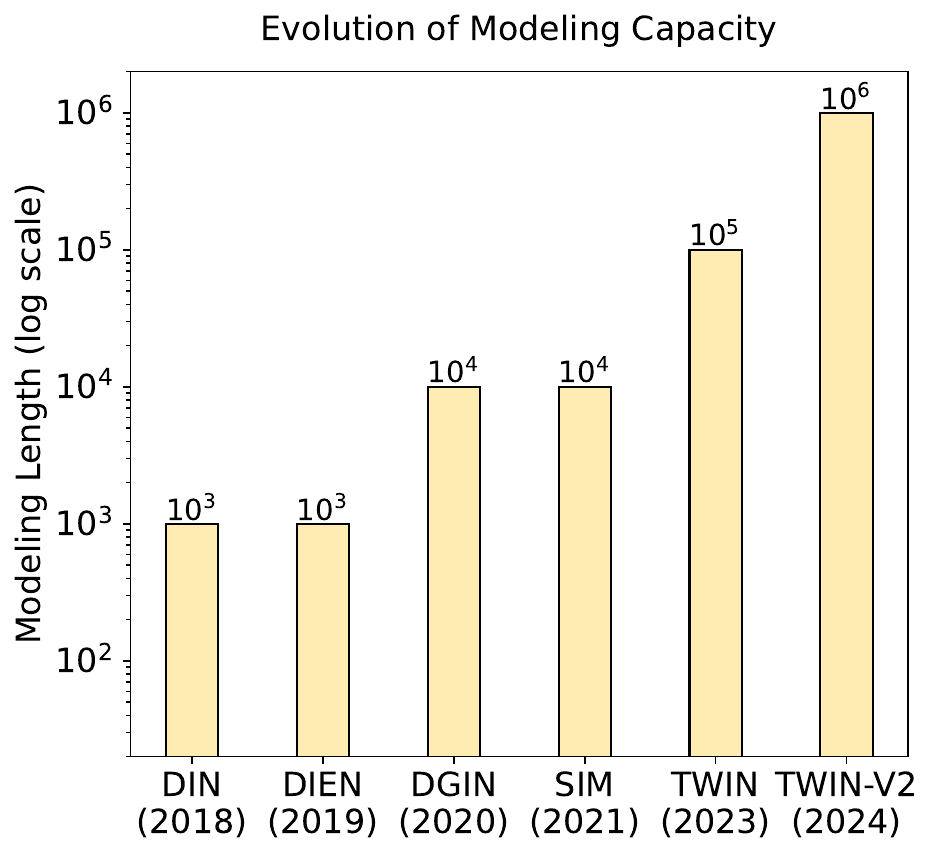}}
    \centerline{(a) Model Capacity Evolution}
  \end{minipage}
  \hfill
  \begin{minipage}[t]{0.49\linewidth}
    \centering
    \raisebox{-0.3mm}{%
      \fbox{\includegraphics[width=\linewidth]{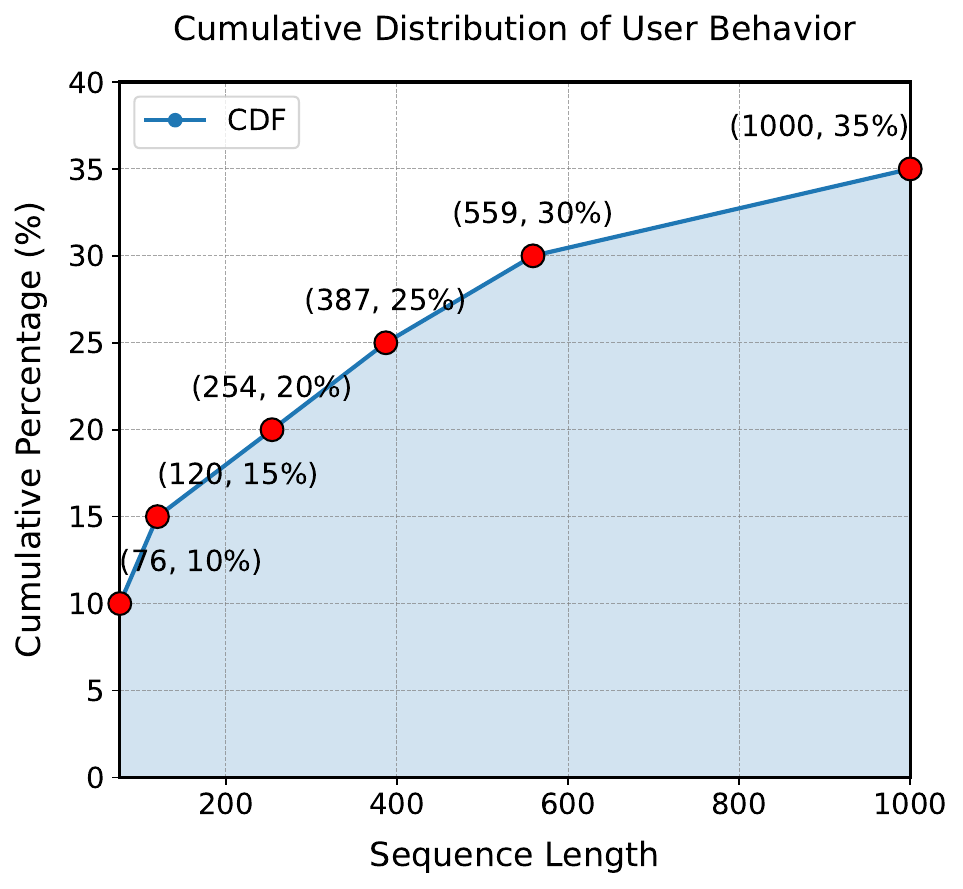}}%
    }
    \centerline{(b) User Behavior CDF}
  \end{minipage}
  \caption{(a) Evolution of CTR model sequence‑length capacity; (b) Cumulative distribution of user sequence lengths over a one‑month period on Kuaishou approximately 35\% of users under 1,000 interactions.}
  \label{figure1}
\end{figure}

Recent research has increasingly focused on modeling long-term user behavior to better capture users’ persistent interests and improve CTR prediction accuracy \cite{liu2023deep,chen2021endtoenduserbehaviorretrieval,cao2022samplingneedmodelinglongterm,liu2024deepgroupmodelinglifelong,zhang2022clustering}. For instance, models like TWIN \cite{chang2023twin} and SIM \cite{pi2020search} employ two-stage frameworks to efficiently handle ultra-long behavior sequences. Building upon this, TWIN-V2 \cite{si2024twin} introduces behavior clustering strategies to capture more precise and diverse user interests. These approaches substantially improve the capacity for long-range sequential modeling and have demonstrated strong performance on industrial datasets and in real-world online deployments. As illustrated in Figure~\ref{figure1}a, the modeling length of CTR models has substantially increased in recent years, expanding from $10^3$ to $10^6$.

Although CTR models have continuously improved in their ability to model long user behavior sequences, such advancements largely rely on data-rich environments. In low-active user scenarios, the lack of sufficient behavioral signals makes it challenging to accurately capture user interests. Even when models can handle long sequences, the scarcity of data leads to a significant decline in prediction performance. As shown in Figure \ref{figure1}b, on the Kuaishou platform, approximately 35\% of users have behavior sequence lengths of fewer than 1,000 within a one-month period, highlighting the real-world sparsity of user behavior data.
Existing data augmentation techniques~\cite{dang2024repeated,liu2023diffusion,liu2021contrastiveselfsupervisedsequentialrecommendation,xie2022contrastive} mainly rely on collaborative signals. 
However, for low-active users, these collaborative signals are inherently sparse and noisy, making it difficult to estimate reliable user similarity or generate semantically meaningful supplementary behaviors. 

Multimodal recommendation methods \cite{liu2024multimodalrecommendersystemssurvey} have shown that textual and visual content can effectively enrich item and user representations. 
Nevertheless, most existing methods incorporate multimodal features as auxiliary inputs to the prediction model, improving the representation of item rather than explicitly providing additional behavioral evidence for low-active users with sparse histories.
Therefore, a natural question arises: can we transform rich item-side multimodal semantics into semantic anchors for low-active users, thereby discovering additional behavior evidence that is consistent with their latent interests?

Motivated by this question, we propose \textbf{MARS} (\textbf{M}odality-\textbf{A}ligned \textbf{R}etrieval for \textbf{S}equence Augmented CTR Prediction). 
Different from existing multimodal recommendation methods that mainly use multimodal features to enrich item representations, MARS transforms item-side multimodal semantics into an aligned retrieval space for retrieving relevant supplementary behaviors for low-active users.
Specifically, MARS first adopts a Stein kernel-based alignment module to calibrate textual and visual item features into a unified semantic space, and then combines them with collaborative signal to construct modality-aligned user representations. 
Based on these representations, MARS retrieves high-active users that are semantically similar to a target low-active user. 
To avoid noisy behavior injection, MARS further applies item-user similarity filtering and only retains behaviors that are relevant to the target user before incorporating them into the sparse user history. 
The proposed method is model-agnostic, MARS operates at the data level and does not require any changes to the CTR model itself.


To summarize, we highlight the key contributions of this paper as follows:
\begin{itemize}[leftmargin=*, itemsep=0pt, topsep=0pt]
\item \textbf{Framework}: We propose MARS, a modality-aligned retrieval framework for sequence-augmented CTR prediction under low-active user sparsity. Different from model-level augmentation or direct multimodal feature fusion, MARS performs data-level behavior augmentation by retrieving and filtering transferable behaviors from semantically similar high-active users while keeping the downstream CTR backbone unchanged.
\item \textbf{Methodology:} 
We introduce a Stein-based multimodal alignment module to construct a retrieval-oriented semantic space. 
The alignment objective balances cross-modal consistency and representation dispersion, leading to more stable multimodal representations for similar-user retrieval and item-user similarity filtering. 
\item \textbf{Experiment:} Extensive offline and online A/B experiments validate that MARS consistently outperforms state-of-the-art methods and improves key business metrics for low-active users. The framework has been deployed, supporting core traffic of Kuaishou for hundreds of millions of users with negligible online overhead.
\end{itemize}

\section{Problem Formulation}

In this section, we will briefly introduce the problem definition used in this paper, in order to aid in the reader’s understanding.
\subsection{Problem Formulation}\label{subsec:hyperbolic_geometry}
CTR prediction is formulated as a supervised binary classification problem within the machine learning framework. Let $\mathcal{D} = \{(u_j, v_j, c_j, l_j)\}_{j=1}^{|\mathcal{D}|}$ denote the training dataset, where each sample comprises a user feature vector $u_j \in \mathbb{R}^{d_u}$, an item feature vector $v_j \in \mathbb{R}^{d_v}$, a contextual feature vector $c_j \in \mathbb{R}^{d_c}$, and the corresponding binary click label $l_j \in \{0,1\}$. The prediction process can be mathematically expressed as:
\begin{equation}
\hat{p}_j = \sigma(\mathcal{M}(u_j, v_j, c_j; \Theta)),
\end{equation}
where $\mathcal{M}(\cdot)$ represents the CTR prediction model parameterized by $\Theta$, which maps the multi-modal feature representations to a real-valued output, and $\sigma(\cdot)$ denotes the sigmoid function that transforms the output into a probability within the interval $(0,1)$.

The model parameters are optimized by minimizing the following cross-entropy loss function:
\begin{equation}
\mathcal{J}(\Theta) = -\frac{1}{|\mathcal{D}|} \sum_{j=1}^{|\mathcal{D}|} \left[ l_j \log(\hat{p}_j) + (1-l_j) \log(1-\hat{p}_j) \right] + \lambda \mathcal{R}(\Theta),
\end{equation}
where $\mathcal{R}(\Theta)$ represents the regularization term and $\lambda$ denotes the regularization coefficient, which collectively prevent overfitting and enhance the model's generalization capability.

\section{Framework}
\begin{figure}
\centering
\includegraphics[width=1\linewidth]{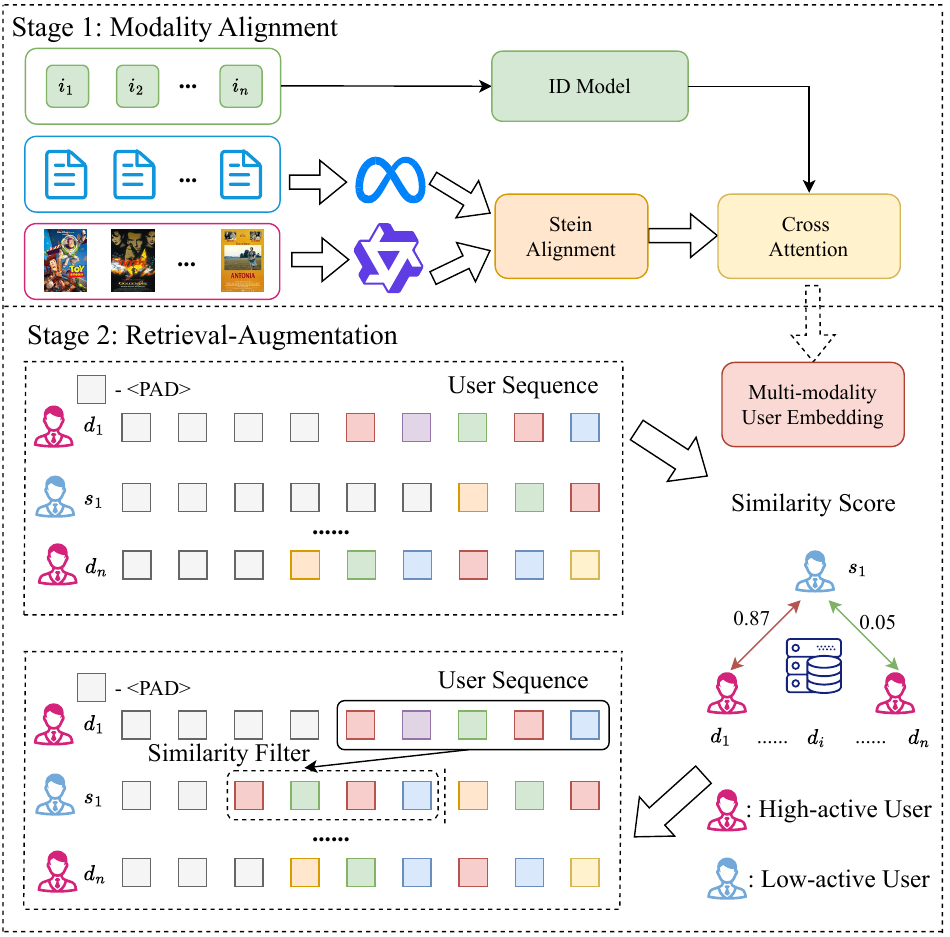}
\caption{An overview illustration of the MARS architecture.}
\label{mars}
\end{figure}

In this section, we introduce MARS, a novel data augmentation framework. As illustrated in Figure \ref{mars}, MARS operates in two stages. First, MARS employs a Stein kernel-based method to align key features extracted from item IDs, text, and images, thereby constructing a unified multimodal user embedding. Subsequently, we propose a simple yet effective data augmentation strategy: for each low-active user, high-active users with similar multimodal embeddings are retrieved. To ensure relevance, the retrieved behavior sequences are further filtered 
based on item-user similarity before being incorporated into the target user’s history.

\subsection{Multimodal Feature Extraction}
Traditional data augmentation methods for recommender systems exhibit significant limitations. These methods primarily rely on simple sequence operations such as random masking, reordering, and cropping, but they face several critical challenges: (1) \textbf{Signal singularity}---they rely solely on collaborative signals from user-item interactions, neglecting the rich multimodal content features inherent in items. Failing to leverage semantic information, such as textual descriptions and visual content to guide the augmentation process; (2) \textbf{Semantic blindness}---lacking a deep understanding of semantic similarity and content associations between items, leading to augmentation operations with random and semantically inconsistent substitutions or transformations; (3) \textbf{Context neglect}---failing to adequately consider sequential temporal context and the evolution of user interests, resulting in augmented samples that may violate users’ actual behavioral logic and preference transitions. These limitations hinder the ability of traditional augmentation methods to generate high-quality, semantically consistent training samples, thereby impairing model generalization and overall recommendation performance.

To address these fundamental limitations,  we design a  multimodal feature extraction module. Specifically, we extract and project modality-specific features for each item as follows:
\begin{equation}
\begin{aligned}
e_{id} &= \text{ItemEmbedding}(v) \\
e_{txt} &= W_{\text{txt}} \cdot f_{\text{LLaMA}}(\text{txt}) \\
e_{img} &= W_{\text{img}} \cdot f_{\text{ViT}}(\text{img}) \\
\end{aligned}
\label{eq:embeddings} 
\end{equation}
where $e_{id}$ encodes unique item identifiers and collaborative information, $e_{txt}$ utilizes large language models to encode item semantic descriptions and attribute information, and $e_{img}$ extracts visual features and appearance information through Vision Transformers.
To capture contextual dependencies among items, we apply a self-attention mechanism over the item embedding sequence $L(e) = [e_{id_1}, e_{id_2}, \dots, e_{id_n}]$ producing a sequence-level representation \(emb_{id} \). Specifically, we compute \( emb_{id} \) as:
\begin{equation}
emb_{id} = \text{softmax}\left( \frac{(L(e)W_Q)(L(e)W_K)^\top}{\sqrt{d_k}} \right)(L(e)W_V),
\end{equation}
where \( W_Q, W_K, W_V \in \mathbb{R}^{d \times d_k} \) are learnable projection matrices for queries, keys, and values, and \( d_k \) is the dimension of the key vectors. This attention mechanism allows the model to capture item dependencies and contextual relevance within the sequence.
\subsection{Stein-Based Multimodal Alignment}
Prior to being fused with \(emb_{id} \), multimodal signals must undergo precise calibration and alignment. MARS achieves this by integrating Stein kernel-based similarity estimation with entropy regularization, ensuring that the embeddings from both modalities remain semantically consistent and information-rich throughout the recommendation pipeline. Compared to traditional alignment methods that rely on negative sampling or large batch training, the Stein approach enables direct distribution alignment without the need for negative pairs, resulting in more stable and efficient training. Importantly, this approach is designed to prevent representation collapse, a common failure mode in vanilla alignment where all embeddings converge to a single point (see Section VI-A for the theoretical proof). 

Specifically, let the image and text modalities be independently encoded, producing final embedding vectors $emb_{img}$ and $emb_{txt}$, respectively. We employ a Radial Basis Function (RBF) kernel \cite{buhmann2000radial} to construct a nonlinear similarity metric defined as:
\begin{equation}
K(emb_{img}, emb_{txt}) = \exp \left( - \frac{\|emb_{img} - emb_{txt}\|^2}{2\gamma^2} \right).
\end{equation}
In implementation, the RBF kernel is computed within each mini-batch over item-level textual and visual embeddings. The bandwidth parameter \(\gamma\) is not manually tuned for each dataset; instead, it is adaptively determined by the median trick, i.e., the median of pairwise cross-modal embedding distances in the current mini-batch. This batch-wise adaptive bandwidth makes the kernel scale consistent with the current embedding distribution and avoids dataset-specific kernel tuning.

This kernel captures higher-order correlations, significantly enhancing the nonlinear alignment capability between modalities. The choice of the RBF kernel is theoretically motivated by its “Locality of Repulsion” property. Unlike linear or polynomial kernels, the RBF kernel’s gradient provides a distance-aware repulsive force that decays exponentially, ensuring that it only prevents collapse when embeddings are excessively close without disrupting the global semantic structure. A detailed comparative justification of the RBF kernel choice is presented in Section VI-B. Based on this, we define the cross-modal consistency loss as:
\begin{equation}
    L_{\mathrm{align}} = \mathbb{E}_{(emb_{img}, emb_{txt})} \left[ K(emb_{img}, emb_{txt}) \right].
\end{equation}
This term is computed on matched image-text pairs from the same item and acts as a cross-modal attraction force. 
Since the overall training objective is minimized, \(-\mathcal{L}_{align}\) is used to maximize the paired cross-modal kernel similarity.

However, naive modality alignment can cause excessive contraction of the embedding space, leading to representation collapse or modality dominance issues \cite{wang2025findrecsteinguidedentropicflow}. 
To mitigate these problems, we introduce an entropy regularization term derived from the approximate Stein score function to enhance the diversity and uniformity of the embedding distribution~\cite{liu2019steinvariationalgradientdescent,wen2024mvebselfsupervisedlearningmultiview}. 
Concretely, we leverage the Stein gradient estimator to approximate the score function \(\nabla_{e}\log q(e)\) of the implicit embedding distribution \(q(e)\) over the current mini-batch, without explicitly estimating the density. 
This score-based approximation serves as a tractable surrogate for the gradient of differential entropy in high-dimensional embedding spaces. 
As mathematically derived in Section~VI-C, the resulting regularizer introduces a kernel-induced local repulsive effect that balances the alignment driving force: when embeddings become overly concentrated, the repulsion becomes stronger, while its effect naturally decays for already separated embeddings due to the RBF kernel. 
Based on this, we formulate the entropy regularizer as:
\begin{equation}
    L_{\mathrm{entropy}} = - \mathbb{E}_{emb \in \{emb_{img}, emb_{txt}\}} \left[ \nabla_{e} \log q(e)^\top e \right].
\end{equation}

Ultimately, the overall loss for the modality alignment module is jointly formulated as:
\begin{equation}
    L_{\mathrm{stein}} = -L_{\mathrm{align}} + \lambda \cdot L_{\mathrm{entropy}},
    \label{stein_gs}
\end{equation}
where $\lambda$ is a balancing hyperparameter controlling the trade-off
between alignment fidelity and information diversity. Through this
dual-objective optimization, high-quality fusion of modalities in
the representation space is achieved, ensuring that the resulting embeddings delivered to downstream tasks maintain both alignment
and fine-grained modality-specific features, thereby significantly
improving the generalization and discriminative power of the recommendation system.

\subsection{Multimodal Fusion and Training Objective}
After aligning the image ($emb_{img}$) and text ($emb_{txt}$) embeddings into a unified semantic space, we perform a dynamic fusion. We employ a cross-modal attention mechanism where the image embedding acts as a “query” to selectively aggregate information from the text embedding, which serves as “keys” and “values”. This allows the model to dynamically determine which parts of the text are most relevant to the visual content. The core operation produces an attention-weighted text representation, $emb_{attn}$, as follows:
\begin{equation}
    emb_{attn} = \text{CrossAttention}(emb_{img}, emb_{txt}).
\end{equation}
The final fused multimodal representation $emb_{fused}$ is obtained by integrating this attended information with the original image embedding via a residual connection: 
\begin{equation}
emb_{fused} = \text{LayerNorm}(emb_{img} + emb_{attn}).
\end{equation}
Next, we apply a secondary cross-attention between \(emb_{fused}\) (as query) and the collaborative embedding \(emb_{id}\) (as keys and values):
\begin{equation}
    emb_{item}^{mm} = \text{CrossAttention}(emb_{\text{fused}}, emb_{\text{id}})
\end{equation}
The predicted score $S$ is then obtained by feeding $emb_{item}^{mm}$ into a feed-forward network to fit the groud truth label:
\begin{equation}
S = \text{FFN}(emb_{item}^{mm})
\end{equation}
The user's multimodal representation $emb_{user}^{mm}$ is obtained by average pooling over the fused multimodal embeddings $emb_{item}^{mm}$ of historical items. Formally, given the set of historical items $\mathcal{H}_{user} = \{item_1, item_2, \ldots, item_N\}$, where each item has a fused multimodal embedding $emb_{item_i}^{mm}$, the user's multimodal representation is computed as:
\begin{equation}
emb_{user}^{mm} = \frac{1}{N} \sum_{i=1}^{N} emb_{item_i}^{mm}
\end{equation}
The entire network of the modality alignment process is trained end-to-end. The final training objective jointly optimizes the primary binary cross-entropy loss for CTR prediction ($\mathcal{L}_{\text{CTR}}$) and the Stein-based alignment loss ($\mathcal{L}_{\text{stein}}$) from Equation \ref{stein_gs} as a regularization term:
\begin{equation}
  \mathcal{L}_{\text{total}}
  = \mathcal{L}_{\text{CTR}}
    + \beta \cdot \mathcal{L}_{\text{stein}},
\end{equation}
where $\beta$ is a balancing hyperparameter. This joint optimization ensures the learned embeddings are predictive and well-aligned, providing a foundation for retrieval augmentation.
\subsection{Retrieval-Augmentation}
To alleviate the data sparsity inherent in low‑active users, we propose a retrieval‑augmentation strategy grounded in aligned multimodal user embeddings. For a given low‑active user \(u_{low}\), we retrieve the most similar high‑active user \(u^*\) from the high-active user set $\mathcal{U}_{high}$ by
\begin{equation}
u^* \;=\;\arg\max_{u \,\in\, \mathcal{U}_{high}}
\frac{\left(emb_{u}^{mm} \right)^\top
  emb_{u_{low}}^{mm}
}
{\|emb_{u}^{mm}\|\|emb_{u_{low}}^{mm}\|}.
\end{equation}
Let the historical behavior sequence of \(u^*\) be $S' = [\,v'_1, v'_2, \dots, v'_n\,]$.
Next, for each item \(v'_i\in S'\), we compute its similarity to \(u_{low}\):
\begin{equation}
s_i \;=\;\frac{\left(emb_{v'_i}^{mm} \right)^\top emb_{u_{low}}^{mm}}
{\|emb_{v'_i}^{mm}\|\|emb_{u_{low}}^{mm}\|},
\end{equation}
and filter by a threshold \(\theta\), retaining only those items with \(s_i \ge \theta\):
\begin{equation}
\widetilde{S}' = [\,v'_i \mid s_i \ge \theta,\;i=1,\dots,n\,].
\end{equation}
Finally, we prepend the filtered, high similarity sequence \(\widetilde{S}'\) to the original sequence of the low‑active user \(S_{u_{low}}=[v_1,\dots,v_m]\), yielding the augmented sequence
\begin{equation}
S_{u_{low}}^{\mathrm{aug}} \;=\;\widetilde{S}'\;\Vert\;S_{u_{low}}.
\end{equation}
This retrieval‑augmentation process injects the most relevant high quality behaviors into the sparse history of low‑active users, thereby enriching their behavioral sequence and improving the accuracy of downstream recommendation models.

\subsection{Deployment of MARS}
\begin{figure}[ht]
    \centering
\includegraphics[width=1.0\linewidth]{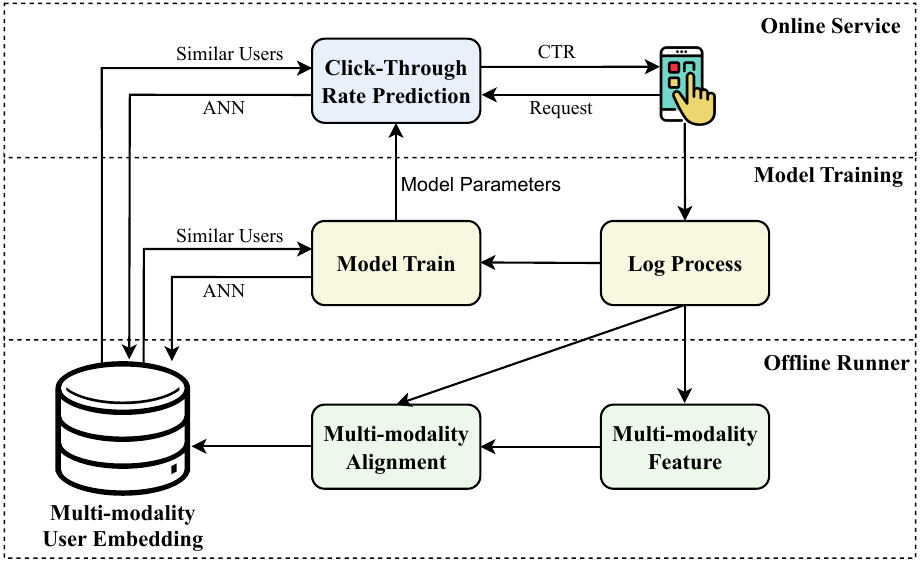}
    \caption{The deployment of MARS at Kuaishou.}
    \label{fig:deployment}
\end{figure}

We divide our hands-on practices for deploying MARS into three parts: \textbf{Model Training}, \textbf{Online Service} and \textbf{Offline Runner}, as shown in Figure~\ref{fig:deployment}.
\subsubsection{Model Training.}The system ingests interaction logs in real time via a message queue, processes the data, and feeds it into model training. To address data sparsity, the training process leverages approximate nearest neighbor (ANN) search in the aligned multimodal semantic space to retrieve semantically similar high-active users, whose behavior sequences augment those of low-active users. Model parameters are synchronized to the online inference service every 20 minutes through the message queue, ensuring that recommendations remain responsive to evolving user interests.
\subsubsection{Offline Runner.}The offline inferring module periodically processes user interaction logs and item metadata to generate multimodal user representations for efficient online lookup. These representations are written to a distributed key–value store and refreshed at minute-level intervals to maintain data freshness while controlling system overhead. Notably, the generated multimodal user embeddings are not limited to use within MARS but can be broadly applied across various downstream tasks, significantly enhancing the overall cost-effectiveness of the pretraining stage.
\subsubsection{Online Service.}Upon receiving a CTR estimation request, the online serving layer first loads the target user’s multimodal embeddings and the item ID embeddings from the offline store. It then performs an ANN search in the aligned embedding space to retrieve behavior sequences of highly active users who are most similar to the target user.
These retrieved sequences are concatenated with the target user’s original behavior trajectory to form an augmented input, which is then fed into a real-time CTR model for low-latency click-through rate prediction.
\section{Algorithmic Workflow}
\label{sec:workflow}
MARS addresses interaction sparsity through a three-stage workflow as summarized in Algorithm \ref{alg:mars}.

\textbf{1) Modality Alignment:} 
Item textual and visual features are first encoded via LLaMA and ViT. We optimize the joint objective $\mathcal{L}_{total}$ (Eq. 14), which integrates the Stein kernel-based alignment loss (Eq. 8). This stage serves as a pre-training phase to ensure the multimodal embeddings ($emb^{mm}$) are both semantically aligned and predictive, constructing the semantic foundation for the subsequent retrieval.

\textbf{2) Retrieval-Augmentation:} 
For each low-active user, we retrieve the most similar high-active user $u^*$ based on the aligned $emb^{mm}$ (Eq. 15). To ensure relevance, retrieved behaviors are refined via similarity filtering (Eq. 16–17) and concatenated with the target user's original history to form the augmented sequence $S^{aug}$ (Eq. 18).

\textbf{3) Final CTR Training:} 
The enriched sequences $\{S^{aug}\}$ are fed into the downstream CTR model $\mathcal{M}$. The final model parameters $\Theta$ are optimized by minimizing the standard cross-entropy loss $J(\Theta)$ defined in Eq. (2). This ensures that the final prediction effectively leverages the injected high-quality behavioral signals, rather than being constrained by the initial alignment objective.

\begin{algorithm}[t]
\caption{MARS Framework Workflow}
\label{alg:mars}
\begin{algorithmic}[1]
\renewcommand{\algorithmicrequire}{\textbf{Input:}}
\renewcommand{\algorithmicensure}{\textbf{Output:}}

\Require 
Training set $\mathcal{D}$; High-active set $U_{high}$; Threshold $\theta$; Hyperparameters $\lambda, \beta$.
\Ensure 
Final optimized CTR model parameters $\Theta$.

\Statex \textit{// Stage 1: Modality Alignment \& Representation Pre-training}
\State Extract item features $e_{txt}$ and $e_{img}$ via LLaMA and ViT
\While{not converged}
    \State Compute Stein-based alignment loss $\mathcal{L}_{stein}$ (Eq. 8)
    \State \textbf{Optimize Alignment Objective:} $\mathcal{L}_{total} = \mathcal{L}_{CTR} + \beta \cdot \mathcal{L}_{stein}$ \Comment{Eq. 14}
    \State \textit{// This stage ensures $emb^{mm}$ is aligned and predictive}
\EndWhile

\Statex \textit{// Stage 2: Retrieval-Based Data Augmentation}
\For{each low-active user $u_{low} \in \mathcal{D}$}
    \State Retrieve similar user $u^*$ based on $emb^{mm}$ (Eq. 15)
    \State Apply similarity filter and extract $S'$ (Eq. 16-17)
    \State Construct augmented sequence $S_{ulow}^{aug} = S' \parallel S_{ulow}$ \Comment{Eq. 18}
\EndFor

\Statex \textit{// Stage 3: Sequence-Augmented CTR Prediction}
\State Feed $\{S^{aug}\}$ into the downstream CTR model $\mathcal{M}$
\State \textbf{Train Final Model:} Minimize standard CTR loss $J(\Theta)$ \Comment{Eq. 2}

\State \Return Optimized parameters $\Theta$
\end{algorithmic}
\end{algorithm}

\section{Experiments}
To evaluate the effectiveness of the proposed MARS framework, we conduct a comprehensive set of experiments on both public datasets and an industrial platform, and we analyze the results with respect to the following research questions:
\begin{itemize}[leftmargin=*]
\item\textbf{RQ1}: How does MARS perform in offline experiments on real-world datasets compared to baseline models?
\item\textbf{RQ2}: How does each proposed module contribute to the performance?
\item \textbf{RQ3}: How does performance vary under different hyperparameter settings?
\item \textbf{RQ4}: How does MARS compare with alternative multimodal alignment methods?
\item \textbf{RQ5}: How do retrieval quality and filtering design affect the effectiveness of MARS?
\item\textbf{RQ6}: How does user activity level affect the performance of MARS?
\item\textbf{RQ7}: How does MARS perform under different padding strategies?
\item\textbf{RQ8}: How does MARS perform in real online CTR Prediction?
\end{itemize}
\subsection{Experimental Setup}
\subsubsection{Datasets}
We conduct comprehensive experiments on four real-world datasets: MovieLens-1M~\footnote{\url{https://grouplens.org/datasets/movielens/1m/}}, MovieLens-10M~\footnote{\url{https://grouplens.org/datasets/movielens/10m/}}, Amazon-Beauty and -Toys~\footnote{\url{https://jmcauley.ucsd.edu/data/amazon/}}, each combining rating information and multimodal content. Detailed statistics are provided in Table \ref{dataset_sta}. Following prior work \cite{zhou2018deep,zhou2019deep}, we adopt the leave-one-out strategy to partition each user's interactions into training, validation, and test sets.
Across all datasets, interactions rated below 4 are classified as false sample. To extract modality-specific features, we utilize LLaMA-7B \cite{touvron2023llama} for textual representation and the ViT encoder from Qwen2.5-VL \cite{bai2025qwen25vltechnicalreport} for visual features. 

\subsubsection{Evaluation Metrics}
We employ three metrics to evaluate CTR prediction performance: AUC, LogLoss, and Relative Improvement (RelaImpr).

\noindent\textbf{AUC.}
AUC measures the model's ability to distinguish positive and negative samples. 
Following industrial practice, we report impression-weighted AUC:
\begin{equation}
\text{AUC} =
\frac{\sum_{i=1}^{n} \#\text{imp}_i \times \text{AUC}_i}
{\sum_{i=1}^{n} \#\text{imp}_i},
\label{eq:auc}
\end{equation}
where $n$ is the number of users, and $\#\text{imp}_i$ and $\text{AUC}_i$ denote impression count and user-level AUC.

\noindent\textbf{LogLoss.}
LogLoss evaluates the accuracy of predicted probabilities:
\begin{equation}
\text{LogLoss} =
-\frac{1}{N}
\sum_{j=1}^{N} 
\left(
y_j \log(\hat{y}_j)
+ (1-y_j)\log(1-\hat{y}_j)
\right),
\label{eq:logloss}
\end{equation}
where $N$ is the number of interactions.

\noindent\textbf{RelaImpr.}
To measure gains over the baseline, we adopt the RelaImpr metric \cite{zhou2018deep}:
\begin{equation}
    \text{RelaImpr} =
    \left(
    \frac{\text{AUC}(\text{model}) - 0.5}
         {\text{AUC}(\text{base}) - 0.5}
    - 1
    \right)
    \times 100\%.
    \label{eq:relaimpr}
\end{equation}
\begin{table}[h]
\centering
\caption{The statistics of the preprocessed datasets, with “Avg.Len” indicating the average item sequence length.}
\renewcommand{\arraystretch}{1.8}
\setlength{\tabcolsep}{4.5pt}
\begin{tabular}{lcccccc}
\toprule
\textbf{Dataset} & \textbf{\#User} & \textbf{\#Item} & \textbf{\#Inter} & \textbf{\#Avg.Len} & \textbf{\#Sparsity} \\
\midrule
MovieLens-1M & 6,041 & 3,417 & 1,000,210 & 165 & 94.95\% \\
MovieLens-10M & 69,879 & 10,678 & 10,000,054 & 143 & 98.65\% \\
Amazon-Beauty & 4,323 & 2,424 & 60,276 & 13.95 & 99.42\% \\
Amazon-Toys & 19,412 & 11,924 & 167,597 & 8.6 & 99.93\% \\
\bottomrule
\end{tabular}
\label{dataset_sta}
\end{table}
\subsubsection{Baselines}
To ensure a thorough assessment, the baselines are organized into two distinct categories.

\noindent\textbf{First, Heuristic Methods.} The First group comprises commonly used heuristic augmentation strategies, notable for their rule-based design, lack of trainable parameters, and high computational efficiency.
\begin{itemize}[leftmargin=*]
    \item \textbf{Random (Ran)} \cite{liu2023diffusion}: This method applies augmentation by adding randomly selected items from the global item set $v$ into each sequence.
    \item \textbf{RepPad} \cite{dang2024repeated}: This method introduces a repeated padding strategy that replaces traditional zero-padding with repetitions of the original user sequence to better utilize idle input space and improve model performance.
    \item \textbf{Random-seq (Ran-S)} \cite{liu2023diffusion}: Instead of sampling from the full item set, this method randomly selects items from the original sequence to serve as augmentation elements.
    \item \textbf{CMR} \cite{xie2022contrastive}: This method employs three augmentation operations inspired by contrastive learning: \textit{Crop} (selecting a subsequence), \textit{Mask} (hiding some items), and \textit{Reorder} (shuffling a local segment). These operations expose the model to different views of the same behavior trace.
    \item \textbf{CMRSI} \cite{liu2021contrastiveselfsupervisedsequentialrecommendation}: This method enhances the CMR approach through the addition of Substitute and Insert, expanding the augmentation set to five.
\end{itemize}
\noindent\textbf{Second, Trainable Methods.}The second category consists of three trainable data augmentation models that learn augmentation strategies through training.
\begin{itemize}[leftmargin=*]
    \item \textbf{CL4SRec} \cite{xie2022contrastive}: This method introduces contrastive learning to capture user preferences by employing data augmentation techniques to create different views of user behavior sequences and maximizing agreement between augmented views of the same sequence.
    \item \textbf{CoSeRec} \cite{liu2021contrastiveselfsupervisedsequentialrecommendation}: This method introduces informative augmentation techniques (substitute and insert) that leverage item correlations to create robust views for contrastive learning, addressing data sparsity and noise issues in user behavior modeling.
    \item \textbf{DiffuASR} \cite{liu2023diffusion}: This method introduces diffusion models to alleviate data sparsity and long-tail issues by designing a Sequential U-Net for discrete embedding generation and employing guidance strategies to align pseudo interactions with user preferences.
    \item \textbf{LLMSeR} \cite{sun2025llmserenhancingsequentialrecommendation}: This method introduces LLM-driven augmentation to alleviate long-tail sparsity by integrating collaborative semantics and employing adaptive reliability validation.
    \item \textbf{BASRec} \cite{dang2024augmentingsequentialrecommendationbalanced}: This method introduces balanced data augmentation that fuses original and augmented sequence representations through mixup operations to generate diverse yet semantically relevant training samples.
\end{itemize}
\subsubsection{Implementation Details}
During the modality alignment phase, we employ the AdamW optimizer \cite{loshchilov2017decoupled} for model training with a learning rate of 0.001. The batch size is consistently set to 512 throughout both the training and evaluation processes. For item ID embeddings, the embedding dimension is uniformly set to 128 across all four datasets. Regarding multimodal features, image features are extracted using the Qwen2.5-VL \cite{bai2025qwen25vltechnicalreport} and textual features using the LLaMA-7B \cite{touvron2023llama}, with initial dimensions of 3584 and 4096, respectively. These features are subsequently projected into a common hidden dimension of 256. 

During the retrieval-augmentation phase, we again utilize the AdamW optimizer with a learning rate of 0.001. The maximum sequence length is fixed at 200, and the embedding size for all four datasets is uniformly configured to 10. Other hyperparameters are maintained consistently with the original settings presented in their respective studies. We adopt the widely cited DIN model as our base CTR prediction model \cite{zhou2018deep}. Based on historical sequence length, the bottom 30\% of users are defined as low-active users, and the top 30\% as high-active users. We implement all experiments in TensorFlow, utilizing a computational environment equipped with 25 Intel Xeon CPU cores (2.10 GHz) and an NVIDIA RTX 4090 GPU, which provides sufficient resources for large-scale training and hyperparameter optimization.
\subsection{Overall Performance (RQ1)}
\begin{table*}[t]
\centering
\renewcommand{\arraystretch}{2}
\fontfamily{ptm}\selectfont
\resizebox{\textwidth}{!}{
\begin{tabular}{c|ccc|ccc|ccc|ccc}
\hline
\multirow{2}{*}{\textbf{Model}} & \multicolumn{3}{c|}{MovieLens-1M} & \multicolumn{3}{c|}{MovieLens-10M} & \multicolumn{3}{c|}{Amazon-Beauty} & \multicolumn{3}{c}{Amazon-Toys} \\
\cline{2-13}
 & AUC $\uparrow$ & Logloss $\downarrow$ & RelaImpr  & AUC $\uparrow$ & Logloss $\downarrow$ & RelaImpr  & AUC $\uparrow$ & Logloss $\downarrow$ & RelaImpr  & AUC $\uparrow$ & Logloss $\downarrow$ & RelaImpr \\
\hline
BaseModel & 0.7772 & 0.5577 & 0.00\% & 0.7284 & 0.6047 & 0.00\% & 0.6404 & 0.5558 & 0.00\% & 0.6486 & 0.5089 & 0.00\% \\
\midrule
\multicolumn{13}{c}{\textbf{Heuristic Methods}}\\
\midrule
CMRSI \textcolor{blue}{(arXiv2021)} & 0.7798 & 0.5563 & 0.94\% & 0.7292 & 0.6042 & 0.35\% & 0.6467 & 0.5549 & 4.48\% & 0.6497 & 0.5087 & 0.74\% \\
CMR \textcolor{blue}{(ICDE2022)} & 0.7754 & 0.5581 & -0.64\% & 0.7289 & 0.6043 & 0.22\% & 0.6453 & 0.5553 & 3.49\% & 0.6491 & 0.5084 & 0.34\% \\
Random \textcolor{blue}{(CIKM2023)} & 0.7738 & 0.5586 & -1.23\% & 0.7277 & 0.6053 & -0.31\% & 0.6382 & 0.5567 & -1.57\% & 0.6464 & 0.5093 & -1.48\% \\
Random-seq \textcolor{blue}{(CIKM2023)} & 0.7745 & 0.5583 & -0.97\% & 0.7281 & 0.6046 & -0.13\% & 0.6395 & 0.5571 & -0.64\% & 0.6473 & 0.5095 & -0.87\% \\
RepPad \textcolor{blue}{(RecSys2024)} & 0.7711 & 0.5594 & -2.21\% & 0.7269 & 0.6056 & -0.66\% & 0.6434 & 0.5552 & 2.14\% & 0.6503 & 0.5077 & 1.14\% \\
\midrule
\multicolumn{13}{c}{\textbf{Trainable Methods}}\\
\midrule
CoSeRec \textcolor{blue}{(arXiv2021)} & 0.7789 & 0.5569 & 0.61\% & 0.7293 & 0.6039 & 0.39\% & 0.6517 & 0.5497 & 8.05\% & 0.6514 & 0.5082 & 1.88\% \\
CL4SRec \textcolor{blue}{(ICDE2022)} & 0.7783 & 0.5571 & 0.39\% & 0.7287 & 0.6044 & 0.13\% & 0.6508 & 0.5484 & 7.41\% & 0.6521 & 0.5062 & 2.36\% \\
DiffuASR \textcolor{blue}{(CIKM2023)} & 0.7779 & 0.5574 & 0.25\% & 0.7298 & 0.6032 & 0.61\% & 0.6492 & 0.5556 & 6.27\% & 0.6499 & 0.5071 & 0.87\% \\
LLMSeR \textcolor{blue}{(arXiv2025)} & 0.7787 & 0.5563 & 0.54\% & 0.7314 & 0.6028 & 1.31\% & 0.6527 & 0.5486 & 8.76\% & 0.6527 & 0.5059 & 2.76\% \\
BASRec \textcolor{blue}{(AAAI2025)} & \underline{0.7793} & \underline{0.5561} & \underline{0.76\%} & \underline{0.7325} & \underline{0.6024} & \underline{1.80\%} & \underline{0.6548} & \underline{0.5445} & \underline{10.26\%} & \underline{0.6534} & \underline{0.5061} & \underline{3.23\%} \\
\rowcolor{gray!25} \textbf{MARS} & \textbf{0.7813*} & \textbf{0.5555*} & \textbf{1.48\%} & \textbf{0.7356*} & \textbf{0.6016*} & \textbf{3.15\%} & \textbf{0.6597*} & \textbf{0.5440*} & \textbf{13.75\%} & \textbf{0.6567*} & \textbf{0.5053*} & \textbf{5.45\%} \\
\hline
\end{tabular}
}

\caption{Overall performance on four different datasets. Bold numbers indicate the highest scores, while underlined numbers represent the second-best results among all. Adding ``\textbf{{\Large *}}'' denotes the models have statistically significant improvements (i.e., two-sided t-test with $p<0.05$).}

\label{tab:performance}
\end{table*}
Table \ref{tab:performance} reports the comprehensive performance of all the compared baselines across four datasets. Based on these results, the main observations are as follows:
\begin{itemize}[leftmargin=*]
\item Compared to heuristic augmentation methods such as Ran, RepPad, and CMRSI, MARS consistently achieves the best performance across all datasets. Notably, some heuristic methods even underperform compared to the base model, indicating that traditional augmentation strategies based on random perturbations or rule-based reconstructions are prone to introducing semantic noise or redundant information, thereby disrupting the original sequence structure. In contrast, MARS employs a multimodal-aligned retrieval mechanism to select semantically consistent behavioral sequences from high-active users for augmentation. This approach enhances the representation of low-active users without interfering with their original intent, effectively alleviating the modeling bottleneck caused by data sparsity.
\item Trainable augmentation methods typically rely on multi-view generation and contrastive loss design, introducing additional parameters and training objectives. Although these methods demonstrate certain performance improvements, their tightly coupled training processes and high sensitivity to hyperparameters hinder efficient and stable deployment in industrial-scale recommendation systems. In comparison, MARS incorporates only a pretraining module to construct a unified semantic space. Its augmentation strategy does not require any additional training objectives or parameter optimization and can operate independently during inference. MARS consistently outperforms these trainable methods across all datasets, suggesting that, rather than the complexity of training strategies, the quality and semantic relevance of augmented data are the primary drivers of performance improvement.
\item MARS consistently demonstrates superior performance across varying levels of data density and user behavior distributions, highlighting its strong generalization capability across diverse scenarios. This robustness does not stem from increased model complexity or capacity expansion, but rather from its retrieval mechanism that constructs high-quality augmented sequences grounded in semantic consistency. In contrast to conventional augmentation approaches that rely on rule-based perturbations or complex training objectives to refine sequence representations, MARS emphasizes the semantic quality of augmented content itself. This enables effective adaptation to diverse behavioral patterns and distributional heterogeneity, ensuring sustained performance across a wide range of recommendation tasks.
\end{itemize}

\subsection{Ablation Study (RQ2)}
To validate the effectiveness of the proposed MARS framework, we conduct a series of ablation studies by removing four key components: the textual modality (MARS w/o Txt), the visual modality (MARS w/o Img), the Stein-based alignment mechanism (MARS w/o Stein), and the item-user similarity filtering module (MARS w/o Filter). 
Table \ref{ablation} presents the results of our experiments on four datasets, from
which we draw the following significant conclusions:
\begin{itemize}[leftmargin=*]
\item Removing either the textual modality (MARS w/o txt) or the visual modality (MARS w/o img) results in significant performance degradation across all four datasets. These results indicate that textual and visual modalities are complementary and both are essential in constructing semantically consistent user representations. The fusion of multimodal modalities not only enriches the semantic expressiveness of user interests, but also provides more precise and informative signals for downstream behavior sequence retrieval, thereby effectively mitigating the representation deficiency in low-active user scenarios.
\item Furthermore, omitting the Stein kernel-based multimodal alignment strategy (MARS w/o stein), despite utilizing both textual and visual modalities, also causes substantial performance degradation. This finding underscores the necessity of performing fine-grained modality alignment to ensure that semantic representations remain both coherent and well-structured. Without Stein-based alignment, the embedding space suffers from modality discrepancies and distributional collapse, which hinders the effectiveness of subsequent retrieval-based data augmentation.
\item Finally,  Removing the item-user similarity filtering module (MARS w/o Filter) also hurts the performance compared with the full MARS model. This result shows that the retrieved high-active users may contain both useful supplementary behaviors and irrelevant behaviors. Directly injecting the entire retrieved sequence can introduce noisy items that do not match the target low-active user's interests, which weakens the benefit of augmentation. By retaining only behaviors with sufficient relevance to the target user, the filtering module makes the augmented sequence cleaner and more personalized, leading to more reliable CTR prediction.
\end{itemize}

Overall, these ablation results validate the necessity and complementary roles of the key components in MARS. 
Textual and visual modalities provide complementary semantic cues for user representation, Stein-based alignment calibrates them into a coherent multimodal space, and item-user similarity filtering prevents irrelevant retrieved behaviors from being injected into the augmented sequence.

\begin{table}[h]
\centering
\caption{Performance of different design variations across the three datasets. The bolded numbers indicate the most significant decline in performance.}
\label{xiaorong}
\renewcommand{\arraystretch}{1.8}
\setlength{\tabcolsep}{4.5pt}
\begin{tabular}{c|cc|cc|cc}
\toprule
\multirow{2}{*}{Model} & \multicolumn{2}{c|}{ML-1M} & \multicolumn{2}{c|}{Amazon-Beauty} & \multicolumn{2}{c}{Amazon-Toys} \\
\cmidrule(lr){2-3} \cmidrule(lr){4-5} \cmidrule(lr){6-7}
& AUC & RelaImpr & AUC & RelaImpr & AUC & RelaImpr \\
\midrule
MARS & 0.7813 & 0.00\% & 0.6597 & 0.00\% & 0.6567 & 0.00\% \\
w/o txt & \textbf{0.7402} & \textbf{-14.61}\% & \textbf{0.6388} & \textbf{-13.09}\% & 0.6473 & -6.01\% \\
w/o img & 0.7503 & -11.02\% & 0.6397 & -12.52\% & 0.6479 & -5.62\% \\
w/o stein & 0.7565 & -8.81\% & 0.6401 & -12.27\% & \textbf{0.6465} & \textbf{-6.51}\% \\
w/o filter & 0.7806 & -0.25\% & 0.6451 & -9.14\% & 0.6516 & -3.25\% \\
\bottomrule
\end{tabular}
\label{ablation}
\end{table}

\subsection{Hyper-Parameter Analysis (RQ3)}
\begin{figure}[h]
    \centering
    \setlength{\abovecaptionskip}{0pt} 
    \includegraphics[width=0.49\textwidth]
{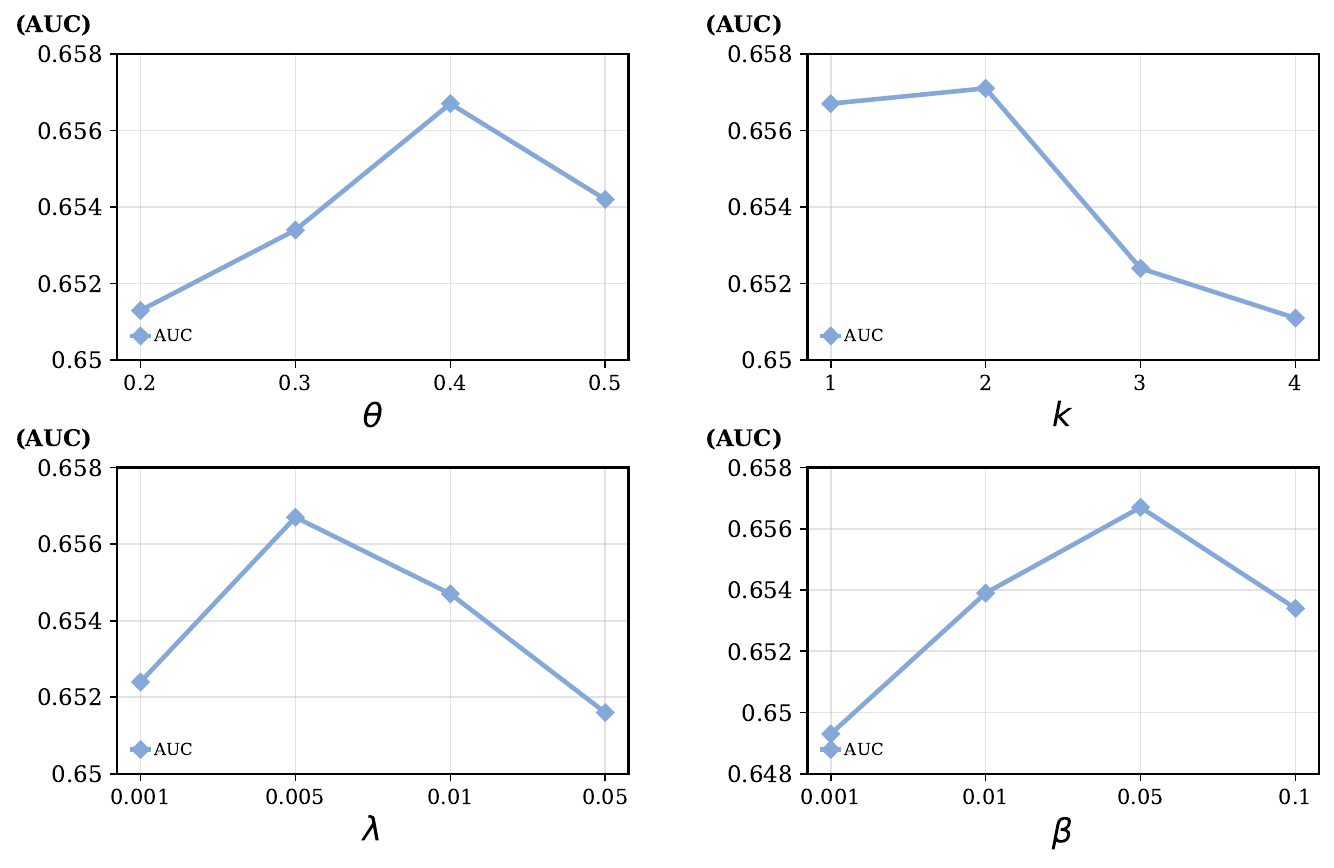}
    \caption{Performance of MARS over different hyper-parameters on Amazon-Toys.}
    \label{chaocan}
\end{figure}

We further investigate several important hyper-parameters introduced in MARS to
facilitate future applications. Specifically, we study the retrieval threshold \(\theta\), the number of retrieved high-active users \(k\),  the strength of entropy regularization  \(\lambda\) and the strength of cross-modal consistency \(\beta\). 
\begin{itemize}[leftmargin=*]
    \item \textbf{Retrieval threshold $\theta$.} 
    We vary the retrieval threshold $\theta$ and report the results in Fig.~\ref{chaocan}. We observe that the model generally achieves a better trade-off with a moderate value of $\theta$. A small $\theta$ preserves many weakly relevant behaviors, which may introduce noise into the augmented sequence. In contrast, an overly large $\theta$ removes many potentially useful behaviors and reduces the coverage of retrieval augmentation.

    \item \textbf{Number of retrieved users $k$.} 
    We further study the effect of $k$ in top-$k$ retrieval while keeping the item-user similarity filtering enabled. The results show that smaller values of $k$ are generally more stable. In particular, $k=1$ or $k=2$ achieves a better balance between retrieval precision and candidate behavior coverage. When $k$ further increases, weakly related high-active users and redundant behaviors may be introduced, which dilutes the personalized signals in the augmented sequence.

    \item \textbf{Strength of Entropy Regularization $\lambda$.} 
    We analyze the sensitivity of $\lambda$ in Fig.~\ref{chaocan}. The results show that a moderate value of $\lambda$ better balances cross-modal semantic consistency and representation dispersion. When $\lambda$ is too small, the entropy-related regularization becomes insufficient, and the embedding space may become over-contracted. When $\lambda$ is too large, the model may over-emphasize representation dispersion, which weakens cross-modal semantic consistency.

    \item \textbf{Strength of Cross-Modal Consistency $\beta$.} 
    We also examine the influence of $\beta$, which controls the contribution of the Stein-based alignment objective. The results indicate that a moderate $\beta$ provides an effective auxiliary constraint for shaping the retrieval space. A small $\beta$ makes the alignment objective too weak to regularize the retrieval representations. In contrast, an overly large $\beta$ may cause the auxiliary alignment objective to dominate the CTR supervision signal, leading to degraded performance.
\end{itemize}

\subsection{Effectiveness of Stein-Based Alignment (RQ4)}
We compare the proposed Stein-based alignment module with two representative alignment alternatives: L2-based alignment and CLIP-style alignment. 
L2-based alignment directly pulls paired image-text embeddings closer, but lacks an explicit mechanism to preserve embedding-space dispersion and may lead to over-contracted representations. 
CLIP-style alignment~\cite{radford2021learning} performs instance-level contrastive learning with in-batch negatives, serving as a strong alignment baseline.

Stein-based alignment is designed to balance cross-modal consistency and representation dispersion. 
Different from L2 alignment, it introduces distribution-aware regularization through the Stein gradient estimator and the RBF kernel, which discourages local over-concentration of embeddings. 
Compared with CLIP-style alignment, it avoids assigning hard negative labels to unpaired samples, which can be problematic in recommendation scenarios where different items may still share categories, visual styles, or user preferences.

Table~\ref{clip} reports the comparison on ML-1M, Amazon-Beauty, and Amazon-Toys. 
Stein-based alignment consistently outperforms the alternatives, indicating that its locality-aware regularization provides a more stable multimodal representation space for retrieval-based sequence augmentation.

\begin{table}[h]
\centering
\caption{Comparison between the proposed Stein-based alignment and other popular alignment method.}
\label{clip}
\renewcommand{\arraystretch}{1.8}
\setlength{\tabcolsep}{4.0pt}
\begin{tabular}{c|cc|cc|cc}
\toprule
\multirow{2}{*}{Method} & \multicolumn{2}{c|}{ML-1M} & \multicolumn{2}{c|}{Amazon-Beauty} & \multicolumn{2}{c}{Amazon-Toys} \\
\cmidrule(lr){2-3} \cmidrule(lr){4-5} \cmidrule(lr){6-7}
& AUC & RelaImpr & AUC & RelaImpr & AUC & RelaImpr \\
\midrule
BaseModel & 0.7772 & 0.00\% & 0.6404 & 0.00\% & 0.6486 & 0.00\% \\
L2 Align & 0.7783 & 0.39\% & 0.6458 & 3.84\% & 0.6493 & 0.47\% \\
CLIP & 0.7791 & 0.68\% & 0.6532 & 9.11\% & 0.6526 & 2.69\% \\
MARS & \textbf{0.7813} & \textbf{1.48\%} & \textbf{0.6597} & \textbf{13.75\%} & \textbf{0.6567} & \textbf{5.45}\% \\
\bottomrule
\end{tabular}
\label{ablation}
\end{table}

\subsection{Retrieval Quality and Setup Analysis (RQ5)}
To further understand the retrieval-based augmentation mechanism in MARS. We analyze this mechanism from two  perspectives: retrieval quality and retrieval setup:

\begin{itemize}[leftmargin=*]
    \item \textbf{Retrieval Quality:} As shown in Table~\ref{jiansuo}, Random Retrieval
(RR) from high-active users  fails to provide consistent gains and even underperforms the Base Model on some datasets. This indicates that simply extending the historical sequences of low-activity users does not effectively alleviate data sparsity and may introduce weakly related or noisy behaviors. In contrast, MARS achieves the best performance across all datasets, suggesting that the gains primarily stem from retrieving high-quality, semantically similar users in the modality-aligned space rather than from sequence length expansion alone.

    \item \textbf{Retrieval Setup:} We further compare MARS with an unfiltered retrieval variant. Filtering the retrieved sequences consistently improves performance, as shown in Table~\ref{jiansuo}. Even if a high-activity user is overall similar to a target low-activity user, their full behavior sequence may contain actions inconsistent with the target’s interests. MARS employs a two-stage retrieval process: first, semantically similar high-activity users are retrieved as candidates; then, item–user similarity filtering retains behaviors that better match the target user’s interests.
\end{itemize}
    These results indicate that the performance improvements arise from the combination of high-quality similar-user retrieval and fine-grained behavior filtering, rather than from simple sequence length extension.
\begin{table}[h]
\centering
\caption{Comparison between the proposed Stein-based alignment and other popular alignment method.}
\label{jiansuo}
\renewcommand{\arraystretch}{1.8}
\setlength{\tabcolsep}{3.5pt}
\begin{tabular}{c|cc|cc|cc}
\toprule
\multirow{2}{*}{Method} & \multicolumn{2}{c|}{ML-1M} & \multicolumn{2}{c|}{Amazon-Beauty} & \multicolumn{2}{c}{Amazon-Toys} \\
\cmidrule(lr){2-3} \cmidrule(lr){4-5} \cmidrule(lr){6-7}
& AUC & RelaImpr & AUC & RelaImpr & AUC & RelaImpr \\
\midrule
BaseModel & 0.7772 & 0.00\% & 0.6404 & 0.00\% & 0.6486 & 0.00\% \\
RR  & 0.7761 & -0.40\% & 0.6388 & -1.13\% & 0.6475 & -0.74\% \\
w/o Filter & 0.7806 & 1.22\% & 0.6451 & 3.35\% & 0.6516 & 2.02\% \\
MARS & \textbf{0.7813} & \textbf{1.48\%} & \textbf{0.6597} & \textbf{13.75\%} & \textbf{0.6567} & \textbf{5.45}\% \\
\bottomrule
\end{tabular}
\label{ablation}
\end{table}

\subsection{User Activity Analysis (RQ6)}
Since MARS targets interaction sparsity, we evaluate its performance under different user activity levels. Users are grouped by training-history length: the bottom 20\% are treated as low-active users and the top 20\% as high-active users. As shown in Table~\ref{user1_gaodihuo}, MARS achieves much larger gains for low-active users. This trend is consistent across datasets, confirming that MARS is especially good at dealing with sparse user segments by supplementing relevant behaviors from semantically similar, high-active users. The small gains for high-active users can be attributed to shared CTR model training, where augmented low-active samples provide richer and more stable training signals.
\begin{table}[h]
\centering
\caption{Performance comparison stratified by user activity 
level. Low-active and high-active groups 
correspond to the bottom 20\% and top 20\% 
of users by sequence length, respectively.}
\label{user1_gaodihuo}
\begin{tabular}{llcc}
\toprule
\textbf{Dataset} & \textbf{User Group} & \textbf{BaseModel} & \textbf{MARS} \\
\midrule
\multirow{4}{*}{MovieLens-1M}
  & Low-active   & 0.7076 & \textbf{0.7311}  \\
  & High-active  & 0.8035 & \textbf{0.8049}  \\
  & Overall      & 0.7772 & \textbf{0.7813}  \\
\midrule
\multirow{4}{*}{Amazon-Beauty}
  & Low-active   & 0.5905 & \textbf{0.6417}  \\
  & High-active  & 0.6927 & \textbf{0.6950}  \\
  & Overall      & 0.6404 & \textbf{0.6597}  \\
\midrule
\multirow{4}{*}{Amazon-Toys}
  & Low-active   & 0.6036 & \textbf{0.6368}  \\
  & High-active  & 0.6973 & \textbf{0.6986} \\
  & Overall      & 0.6486 & \textbf{0.6567}  \\
\bottomrule
\end{tabular}
\end{table}

We further analyze the sensitivity to user-group definitions. 
In Table~\ref{gaodihuo}, \(x\%/x\%\) indicates that the bottom \(x\%\) users by training-history length are augmented using candidates from the top \(x\%\) high-active users, and the table reports the overall AUC under each setting. 
MARS consistently improves over BaseModel across all tested ratios, indicating that its effectiveness is not dependent on a fixed threshold. 
\begin{table}[h]
\centering
\caption{Sensitivity analysis of low-active/high-active user partition ratios.}
\label{gaodihuo}
\renewcommand{\arraystretch}{1.8}
\setlength{\tabcolsep}{3.5pt}
\begin{tabular}{c|cc|cc|cc}
\toprule
\multirow{2}{*}{Method} & \multicolumn{2}{c|}{ML-1M} & \multicolumn{2}{c|}{Amazon-Beauty} & \multicolumn{2}{c}{Amazon-Toys} \\
\cmidrule(lr){2-3} \cmidrule(lr){4-5} \cmidrule(lr){6-7}
& AUC & RelaImpr & AUC & RelaImpr & AUC & RelaImpr \\
\midrule
BaseModel & 0.7772 & 0.00\% & 0.6404 & 0.00\% & 0.6486 & 0.00\% \\
10\%/10\%  & 0.7781 & 0.32\% & 0.6463 & 4.20\% & 0.6504 & 1.21\% \\
20\%/20\% & 0.7795 & 0.83\% & 0.6539 & 9.62\% & 0.6541 & 3.70\% \\
30\%/30\% & \textbf{0.7813} & \textbf{1.48\%} & \textbf{0.6597} & \textbf{13.75\%} & \textbf{0.6567} & \textbf{5.45}\% \\
\bottomrule
\end{tabular}
\label{ablation}
\end{table}

\begin{figure*}[ht]
    \centering
    \setlength{\abovecaptionskip}{0pt}  
    \includegraphics[width=\textwidth]{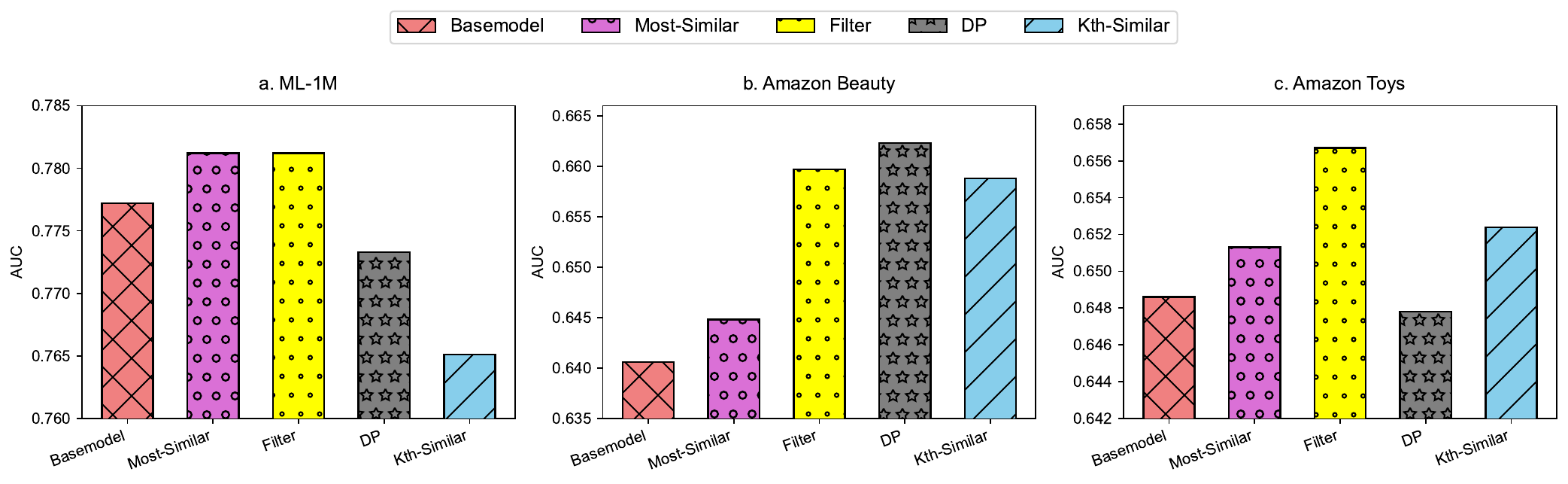}
    \caption{Impact of different padding strategies on model performance (AUC).}
    \label{strategy}
\end{figure*}
\subsection{Padding Strategy (RQ7)}
To investigate the impact of different padding strategies on the performance of MARS, we evaluate the following four methods: (1) \textbf{Most-Similar} – copying the complete interaction sequence from the most similar highly active user to the target user; (2) \textbf{Filtered Most-Similar (Filter)} – an enhanced version of the Most-Similar approach, which filters out low-relevance items based on embedding similarity; (3) \textbf{Kth-Similar} – aggregating interaction sequences from the top-k most similar highly active users; and (4) \textbf{Dynamic Programming (DP)} – utilizing a dynamic programming algorithm to identify the most relevant subsequence from candidate user sequences. According to the experimental results presented in Figure \ref{strategy}, we derive the following three core insights:

\begin{itemize}[leftmargin=*]
\item Across all three datasets ML-1M (dense), Beauty and Toys (sparse) , Most-Similar consistently yields notable performance gains. Filter further refines this approach by removing items with low embedding similarity. This is especially effective in sparse settings, where it filters out long-tail or noisy interactions, leading to greater improvements than direct sequence copying.
\item In the ML-1M dataset, both Kth-Similar and DP introduce excessive popular items and redundant behaviors from similar users, which dilute personalized signals and lead to degraded performance. In contrast, under sparse conditions such as Beauty and Toys, Kth-Similar effectively expands the coverage of long-tail interests by integrating diverse information from multiple similar users, thus compensating for the limitations of single-sequence completion and enhancing recommendation quality.
\item DP achieves SOTA performance on the Beauty dataset, likely due to the highly structured nature of beauty purchase sequences (e.g., “cleanser → base → treatment”), which allows the model to extract rich, reusable subsequences. However, in the Toys dataset, characterized by broad category diversity and highly volatile interests, user behavior is more fragmented, resulting in fewer reusable subsequences. In such cases, DP may introduce noise, thereby undermining recommendation performance.
\end{itemize}

\subsection{Online Result (RQ8)}
To evaluate the online performance of MARS, we conducted a two-week online A/B test on the Kuaishou short video platform. Details of the online experimental protocol and deployment implementation are provided in Appendix A and B. The experimental results demonstrate that our proposed model outperforms the baseline across six key business metrics, achieving statistically significant improvements. As shown in Table \ref{dapan}, although the improvements over the entire user population are modest in magnitude, consistent positive gains are
observed across all metrics. 

Beyond evaluating overall benefits, we further analyzed the model’s specific impact on low-activity users—a critical segment for user growth and ecosystem health. Table \ref{dihuo} shows that the average app usage time per low-activity user increased by an impressive 0.728\%, significantly exceeding the average improvement across all users. Moreover, this group exhibited notable gains in both 7-day usage duration (+0.634\%) and 7-day active hours (+0.310\%) after exposure to the model. These results validate the effectiveness of our approach in uncovering and addressing the latent interests of under-engaged users, successfully converting them into more active participants and positively influencing short-term retention. 

Given that MARS augments low-activity users with behaviors from high-active users, there is a potential risk of injecting mainstream preferences and exacerbating popularity bias at the platform level. To assess this concern, we further analyzed ecosystem-level metrics in our two-week online A/B test on Kuaishou. Specifically, we measured average long-tail video exposure, cold-start video exposure, long-tail video watch time and cold-start video watch time. As reported in Table~\ref{changwei}, the relative lifts are +0.011\%, -0.019\%, +0.023\%, and +0.014\%, respectively, with confidence intervals covering zero. These results indicate that MARS does not reduce exposure to long-tail or cold-start content. 
\begingroup                      
  \renewcommand{\arraystretch}{1.8}
\setlength{\tabcolsep}{5pt}    
  \captionsetup[table]{aboveskip=2pt, belowskip=2pt}

  \begin{table}[htbp]
    \centering
    \caption{Relative improvements for all users.}
    \label{dapan}
    \begin{tabular}{lcc}
      \toprule
\textbf{Metric for All Users} & \textbf{RL (\%)} & \textbf{CI} \\      \midrule
      Avg.\ App Usage Time per Device & +0.059\% & 	[0.00\%, 0.10\%] \\
      Avg.\ App Usage Time per User   & +0.049\% & 	[0.01\%, 0.11\%] \\
      Video Watch Time on Featured Page & +0.099\% & 	[0.03\%, 0.17\%] \\
      \bottomrule
    \end{tabular}
  \end{table}


\begin{table}[h]
\centering
\caption{Relative improvements for low-active users.}
\label{dihuo}
\begin{tabular}{lcc}  
\toprule
\textbf{Metric for Low-Activity Users} & \textbf{RL (\%)} & \textbf{CI} \\  
\midrule
Avg.\ App Usage Time per User         & +0.728\% & 	[0.12\%, 1.43\%] \\
7-Day App Usage Time Post-Enrollment  & +0.634\% & [0.11\%, 1.27\%] \\
7-Day Active Hours Post-Enrollment    & +0.310\% & 	[0.06\%, 0.62\%] \\
\bottomrule
\end{tabular}
\end{table}

\begin{table}[h]
\centering
\caption{Impact of MARS on cold-start and long-tail metrics.}
\label{changwei}
\begin{tabular}{lcc}  
\toprule
\textbf{Metric for cold-start and long-tail} & \textbf{RL (\%)} & \textbf{CI} \\  
\midrule
Avg. Long-tail Video Exposure   & +0.011\% & {[}-0.15\%, 0.17\%{]} \\
Avg. Cold-start Video Exposure      & -0.019\% & {[}-0.20\%, 0.16\%{]} \\
Avg. Long-tail Video Watch Time & +0.023\% & {[}-0.12\%, 0.08\%{]} \\
Avg. Cold-start Video Watch Time & +0.014\% & {[}-0.23\%, 0.21\%{]} \\
\bottomrule
\end{tabular}
\end{table}
\par %
\vspace{3pt}
  \small \textit{Note: RL denotes Relative Lift over the production baseline; CI denotes the 99\% confidence interval.}
\endgroup

\section{Theoretical Analysis of Stein Kernel Alignment}
\label{sec:theoretical_analysis}

In this section, we provide the theoretical justification for the modality alignment module in MARS. We first demonstrate that vanilla alignment objectives lead to representation collapse. Subsequently, we derive how the Stein kernel approach mitigates this via implicit entropy regularization and justify the necessity of the Radial Basis Function (RBF) kernel from a geometric perspective.

\subsection{The Risk of Representation Collapse in vanilla alignment}

Let $\mathcal{D} = \{(\mathbf{x}_v, \mathbf{x}_t)\}$ denote paired image and text inputs, with encoders $f_v(\cdot)$ and $f_t(\cdot)$ producing embeddings $\mathbf{e}_v$ and $\mathbf{e}_t \in \mathbb{R}^d$. A vanilla alignment objective typically minimizes the expected Euclidean distance:
\begin{equation}
\mathcal{L}_{\text{naive}} = \mathbb{E}_{(\mathbf{x}_v, \mathbf{x}_t) \sim \mathcal{D}} \left[ \| \mathbf{e}_v - \mathbf{e}_t \|_2^2 \right]. \label{eq:naive_loss}
\end{equation}

\begin{proposition} \label{prop:collapse}
The objective $\mathcal{L}_{\text{naive}}$ admits a trivial global minimum where encoders map all inputs to a constant vector $\mathbf{c} \in \mathbb{R}^d$.
\end{proposition}

\begin{proof}
By setting $f_v(\mathbf{x}) = f_t(\mathbf{x}) = \mathbf{c}$, Eq. \eqref{eq:naive_loss} yields $\mathcal{L}_{\text{naive}} = \mathbb{E} [ \| \mathbf{c} - \mathbf{c} \|_2^2 ] = 0$. Since the loss is non-negative, this state is a global minimum.
\end{proof}

This phenomenon, termed \textit{representation collapse}, causes the embedding space to contract onto a single point, annihilating the variance of the feature distribution. From an information-theoretic perspective, the model fails to preserve the entropy of the input data, rendering the embeddings non-discriminative. For CTR prediction, this collapse is fatal: it leads to “semantic blindness” where the retrieval mechanism cannot distinguish between distinct user interests, thereby injecting noisy and irrelevant sequences into the augmentation process. Consequently, minimizing Euclidean distance on positive pairs alone is insufficient for learning informative multimodal representations.

\subsection{Justification for the RBF Kernel}

The efficacy of the repulsive force in preventing collapse depends critically on the choice of the kernel $k(\mathbf{e}', \mathbf{e})$. We employ the Radial Basis Function (RBF) kernel, defined as $k(\mathbf{e}', \mathbf{e}) = \exp \left( - \frac{\| \mathbf{e}' - \mathbf{e} \|_2^2}{2\sigma^2} \right)$. The gradient of the RBF kernel with respect to $\mathbf{e}'$ is:
\begin{equation}
\nabla_{\mathbf{e}'}k(\mathbf{e}', \mathbf{e}) = -\frac{1}{\sigma^2} (\mathbf{e}' - \mathbf{e}) \cdot \exp \left( - \frac{\| \mathbf{e}' - \mathbf{e} \|_2^2}{2\sigma^2} \right). \label{eq:rbf_gradient}
\end{equation}

Compared to linear or polynomial kernels, the RBF kernel offers three distinct theoretical advantages in a high-dimensional embedding space:

\begin{itemize}[leftmargin=*]
    \item \textbf{Locality of Repulsion}: As shown in Eq. \eqref{eq:rbf_gradient}, the repulsive force vanishes exponentially as the distance between embeddings increases. This ``short-range" repulsion ensures that force is only exerted when two embeddings are in danger of collapsing. Crucially, it allows distant semantic clusters to evolve independently without being affected by unnecessary global repulsion, thereby preserving the global manifold structure.
    
    \item \textbf{Adaptive Gradient Magnitude}: The term $(\mathbf{e}' - \mathbf{e})$ ensures that the magnitude of the force scales linearly with proximity when embeddings are close, while the exponential term gates the interaction range. This dual mechanism provides a stable ``repulsive buffer" that prevents local crowding while allowing the alignment objective to dominate once a sufficient margin is established.
    
    \item \textbf{Boundedness and Smoothness}: As a stationary and bounded kernel, the RBF kernel produces a Lipschitz continuous gradient field. This property is vital for optimization stability in deep recommendation models, as it prevents the gradient explosion or vanishing issues that often plague non-bounded kernels (like the linear kernel where $\nabla k$ is constant) in sparse high-dimensional spaces.
\end{itemize}


\subsection{Entropy Regularization via Stein Variational Gradient}

To mitigate the representation collapse proven in Proposition \ref{prop:collapse}, MARS aims to optimize the distribution of learned embeddings $q(\mathbf{e})$ such that it aligns with a target distribution $p(\mathbf{e})$ (representing ideal semantic alignment) while maintaining maximum diversity. This is formulated as minimizing the Kullback-Leibler (KL) divergence:
\begin{equation}
\min_{q} \text{KL}(q \| p) = \mathbb{E}_{\mathbf{e} \sim q} [\log q(\mathbf{e}) - \log p(\mathbf{e})]. \label{eq:kl_div}
\end{equation}
Expanding Eq. \eqref{eq:kl_div} reveals a dual-objective structure:
\begin{equation}
\text{KL}(q \| p) = \underbrace{-\mathbb{E}_{\mathbf{e} \sim q}[\log p(\mathbf{e})]}_{\text{Likelihood (Alignment)}} - \underbrace{\mathbb{E}_{\mathbf{e} \sim q}[-\log q(\mathbf{e})]}_{\text{Entropy } H(q)}. \label{eq:kl_expansion}
\end{equation}
Minimizing Eq. \eqref{eq:kl_expansion} is equivalent to maximizing the alignment likelihood while simultaneously maximizing the differential entropy $H(q)$. The latter serves as a natural counterforce to over-contraction, encouraging embeddings to be dispersed across the feature space.

However, direct optimization of $H(q)$ is intractable in high-dimensional spaces as it requires knowledge of the unknown density $q(\mathbf{e})$. To circumvent this, we utilize Stein Variational Gradient Descent (SVGD) \cite{liu2016stein,liu2017stein}. SVGD provides a non-parametric update rule that transforms the embedding distribution through a velocity field $\boldsymbol{\phi}(\mathbf{e})$. In the Reproducing Kernel Hilbert Space (RKHS) $\mathcal{H}_k$, the optimal update direction $\boldsymbol{\phi}^*(\mathbf{e})$ that yields the steepest descent in KL divergence is given by:
\begin{equation}
\boldsymbol{\phi}^*(\mathbf{e}) = \mathbb{E}_{\mathbf{e}' \sim q} \left[ \underbrace{k(\mathbf{e}', \mathbf{e}) \nabla_{\mathbf{e}'} \log p(\mathbf{e}')}_{\text{Driving Force}} + \underbrace{\nabla_{\mathbf{e}'} k(\mathbf{e}', \mathbf{e})}_{\text{Repulsive Force}} \right]. \label{eq:svgd_update}
\end{equation}

This formulation decomposes the alignment process into two geometrically interpretable components:
\begin{itemize}[leftmargin=*]
    \item \textbf{Driving Force}: This term acts as a “semantic attractor,” pulling embeddings toward high-probability regions of $p(\mathbf{e})$ to enforce cross-modal consistency. The kernel $k(\mathbf{e}', \mathbf{e})$ serves as a weighted mask, aggregating alignment signals from neighboring particles.
    \item \textbf{Repulsive Force}: Derived from the Stein operator, this term introduces a “locality-aware repulsion” that pushes embeddings away from each other when they congregate too closely. Unlike contrastive learning, which requires computationally expensive negative sampling, the repulsive force in MARS is computed directly from the kernel gradient, providing a more efficient and stable mechanism for entropy regularization.
\end{itemize}

By integrating Eq. \eqref{eq:svgd_update} into the training pipeline, MARS ensures that the multimodal user embeddings remain both semantically coherent and information-rich, providing a robust foundation for subsequent retrieval-augmentation.

\section{Related Work}
In this section, we review two relevant prior works: Click-through rate model and data augmentation.
\subsection{Click-Through Rate Model}
Research on CTR prediction starts with linear and shallow factorization models, including Logistic Regression (LR) \cite{mcmahan2013ad} Factorization Machines (FM) \cite{rendle2010factorization}, and their field-aware variant, Field-aware Factorization Machines (FFM) \cite{juan2016field}.
With the rise of deep learning, models such as Wide \& Deep \cite{cheng2016wide}, DeepFM \cite{guo2017deepfm}, and DCN \cite{wang2017deep,wang2021dcn} combined the memorization capacity of linear models with the generalization power of deep networks, enabling end-to-end learning of high-order feature interactions. To model evolving user interests, the DIN family progressively advances
behavior-sequence modeling from relevance-aware attention to interest
evolution and unified short- and long-term preference modeling \cite{zhou2018deep,zhou2019deep,feng2019deep,chen2019behavior}.

Recent advances in CTR prediction have largely followed two complementary
directions: scaling the capacity and efficiency of CTR models, and extending
the horizon of user behavior modeling. Along the first direction, RankMixer
and TokenMixer-Large~\cite{zhu2025rankmixer,jiang2026tokenmixer} explore scalable
architectures for large-scale industrial ranking, while QNN-$\alpha$ \cite{li2025revisiting} and
FCN \cite{li2026fcn} improve feature crossing mechanisms to capture
more expressive high-order feature interactions.

Along the second direction,long-term behavior modeling has become essential for
CTR prediction. Memory-based methods such as HPMN~\cite{Ren_2019} and
MIMN~\cite{Pi_2019} retain long-term user interests, while two-stage
retrieval-then-modeling frameworks, including SIM
~\cite{qi2020searchbasedusermodelinglifelong} and UBR4CTR
~\cite{Qin_2020,qin2023learning}, retrieve target-relevant subsequences to
separately model short- and long-term interests. ETA \cite{chen2021endtoenduserbehaviorretrieval} and SDIM \cite{cao2022samplingneedmodelinglongterm} introduce locality-sensitive hashing and signature-based mechanisms to improve retrieval precision. TWIN-V2~\cite{si2024twin} scales historical modeling to $10^6$ behaviors via hierarchical clustering and cluster-aware attention. More recently, MIRRN~\cite{xu2025multi} and UxSID~\cite{zhang2026uxsid} further
advance long-sequence modeling through multi-granularity interest retrieval
and Semantic-ID-guided ultra-long sequence compression, respectively.

Existing methods mainly improve CTR model capacity or exploit the target
user's available history. MARS instead targets low-active users with
insufficient histories by retrieving and filtering behaviors from similar
high-active users based on aligned multimodal representations.

\subsection{Data Augmentation}
Data augmentation has been widely used to mitigate data sparsity in recommendation systems. Existing methods generally fall into two categories: heuristic-based and training-based. Heuristic methods apply predefined operations to perturb user behavior sequences \cite{dang2024repeated,liu2023diffusion,sun2019bert4recsequentialrecommendationbidirectional,xie2022contrastive,liu2021contrastiveselfsupervisedsequentialrecommendation,liu2024multimodalrecommendersystemssurvey,zhou2023comprehensive}. These methods are lightweight and easy to deploy, but the generated data often lack semantic coherence and fail to capture deep user preferences. In contrast, training-based methods utilize learnable modules to generate higher-quality data \cite{xie2022contrastive,liu2021contrastiveselfsupervisedsequentialrecommendation,Qiu_2022,bian2022relevant,liu2023diffusion,dang2025dataaugmentationfreelunch,dang2024augmentingsequentialrecommendationbalanced}. For instance, CL4SRec \cite{xie2022contrastive} and CoSeRec \cite{liu2021contrastiveselfsupervisedsequentialrecommendation} construct contrastive views with heuristic operators and optimize them via self-supervised learning. DuoRec \cite{Qiu_2022} employs contrastive regularization to enhance representation robustness, while ReDA \cite{bian2022relevant} retrieves auxiliary user sequences for representation-level augmentation. BASRec \cite{dang2024augmentingsequentialrecommendationbalanced} further balances relevance and diversity through mixup-based single- and cross-sequence augmentations in the representation space. However, most existing methods focus on
collaborative signals and overlook the potential of multimodal features for interest modeling. To bridge this gap, we propose MARS, which leverages aligned multimodal representations to retrieve and augment semantically consistent user sequences, particularly benefiting low-activity users.
\section{Conclusion}
Motivated by the challenges posed by interaction sparsity in CTR prediction, particularly for low-active users, we set out to explore a data augmentation approach to address this issue. Given the limitations of existing augmentation techniques, we focused on leveraging the rich multimodal features of items for enhancement. To this end, we proposed the MARS framework, whose superiority was demonstrated through extensive online and offline experiments. We believe this work represents a significant step in employing multimodal features for CTR data augmentation and lays a valuable foundation for future research in this area.

\section{Acknowledgements}
This work was supported by the National Key Research and Development Program of China under Grant No. 2024YFF0729003, the National Natural Science Foundation of China under Grant Nos. 62176014, 62276015, and 62206266, and the Fundamental Research Funds for the Central Universities.

\bibliographystyle{ieeetr}

\bibliography{Reference}

\begin{IEEEbiography} [{\includegraphics[width=1in,height=1.25in,clip,keepaspectratio]{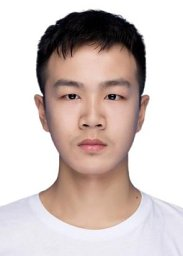}}] {Yutian Xiao} is currently working toward the PhD de-
gree with the Institute of Artificial Intelligence, Bei-
hang University, Beijing, China. His main research
interests include recommender system and natural
language processing.
\end{IEEEbiography}

\begin{IEEEbiography}[{\includegraphics[width=1in,height=1.25in,clip,keepaspectratio]{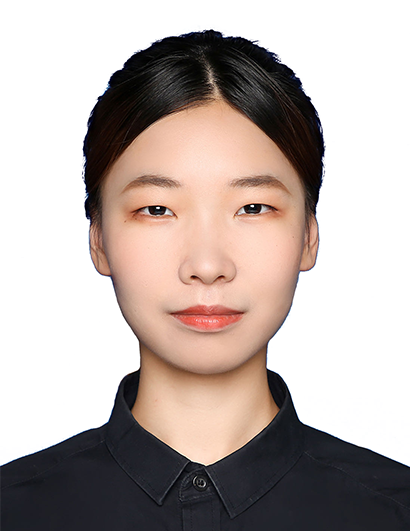}}]{Shukuan Wang}
Currently I am a M.Sc. student of School of Artificial Intelligence in Nanjing University and a member of LAMDA Group, led by professor Zhi-Hua Zhou. I received my B.Sc. degree from School of Artificial Intelligence in June 2023. In September 2023, I was admitted to pursue a M.Sc. degree in Nanjing University, under the supervision of Associate Professor Chao Qian. 
\end{IEEEbiography}

\begin{IEEEbiography}[{\includegraphics[width=1in,height=1.25in,clip,keepaspectratio]{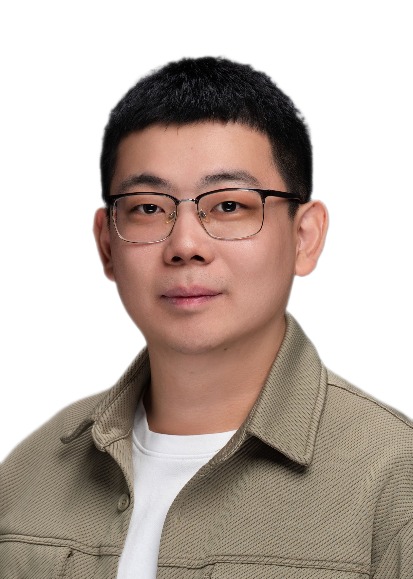}}]{Zhao Zhang} is an associate professor at the school of computer science and engineering, Beihang University. He received the BE degree in computer science and technology from the Beijing Institute of Technology (BIT), Beijing, China, in 2015, and the PhD degree from the Institute of Computing Technology, Chinese Academy of Sciences, Beijing, in 2021. His research interests include data mining and applied machine learning. Zhao has published more than 50 papers in the prestigious refereed journals and conferences like IEEE Transactions on Knowledge and Data Engineering, ACM Transactions on Information Systems, KDD, SIGIR, etc.
\end{IEEEbiography}

\begin{IEEEbiography} [{\includegraphics[width=1in,height=1.25in,clip,keepaspectratio]{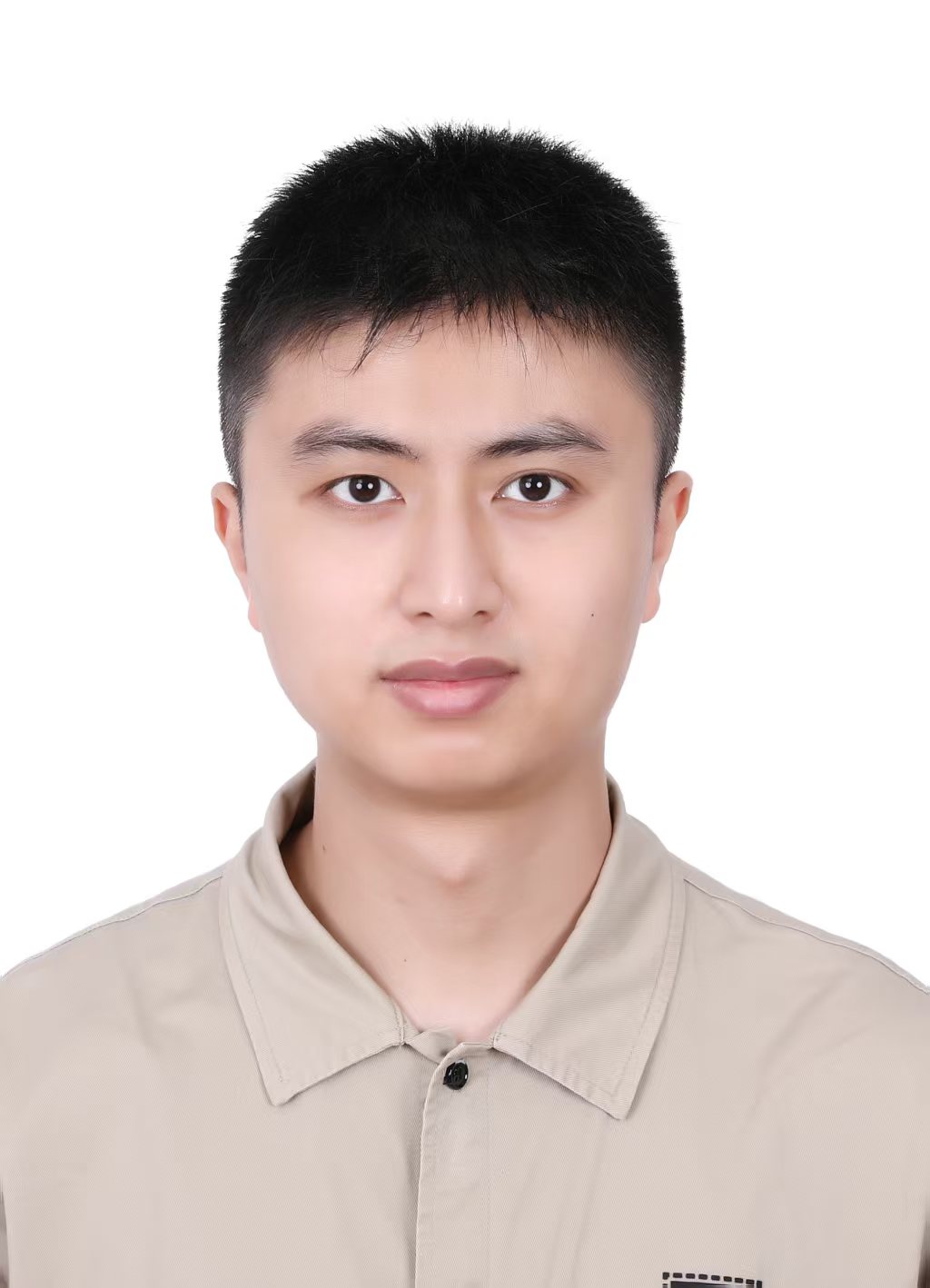}}] {Meng Yuan} is currently pursuing his Ph.D. degree in the Institute of Artificial Intelligence, Beihang University, Beijing, China. His main research interests include recommender system, hyperbolic geometry and machine learning.
\end{IEEEbiography}

\begin{IEEEbiography} [{\includegraphics[width=1in,height=1.25in,clip,keepaspectratio]{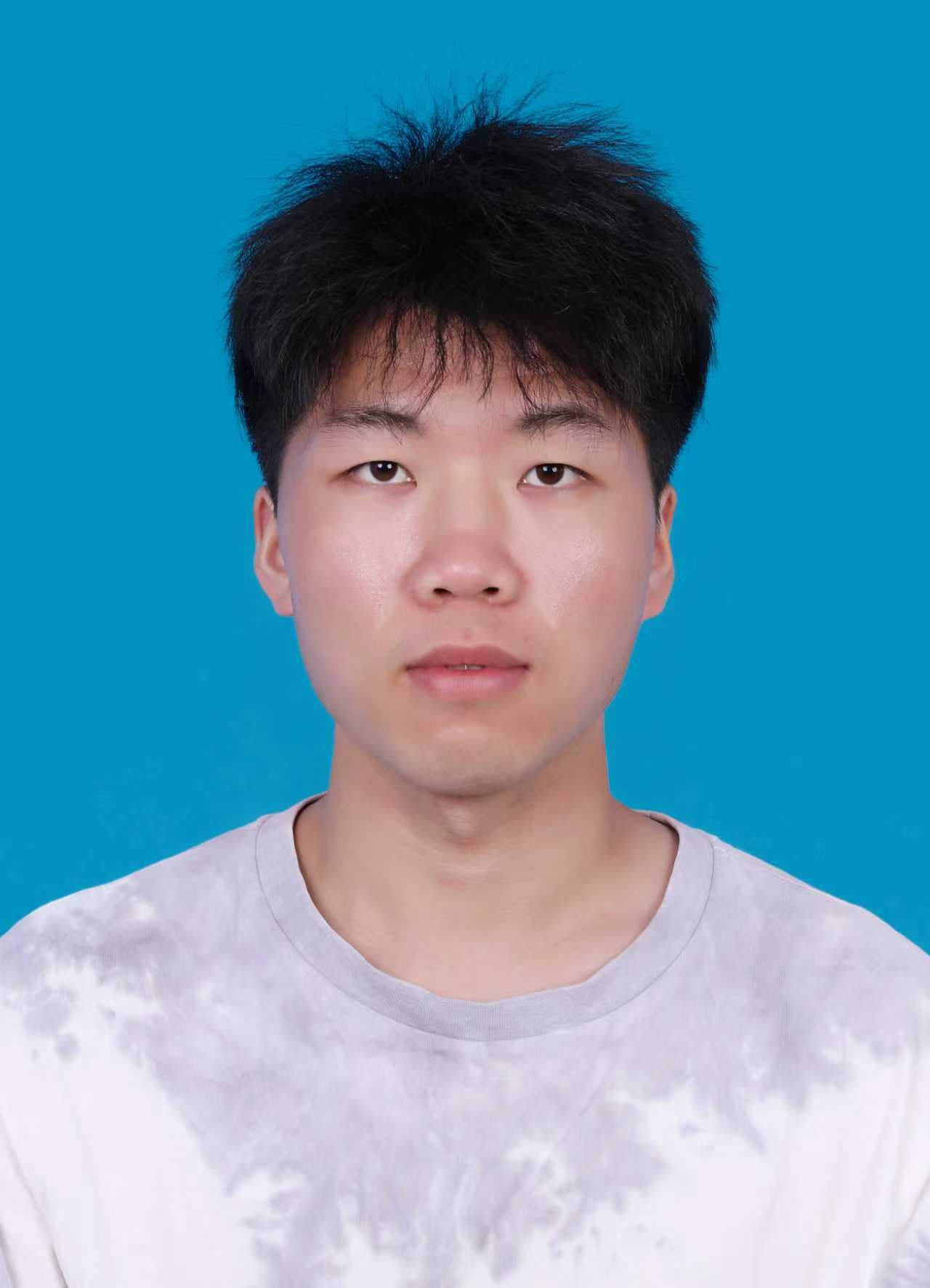}}]{Wei Chen} is currently working toward the PhD degree with the Institute of Artificial Intelligence, Bei- hang University, Beijing, China. His main research interests include recommender system and natural language processing.
\end{IEEEbiography}

\begin{IEEEbiography}[{\includegraphics[width=1in,height=1.25in,clip,keepaspectratio]{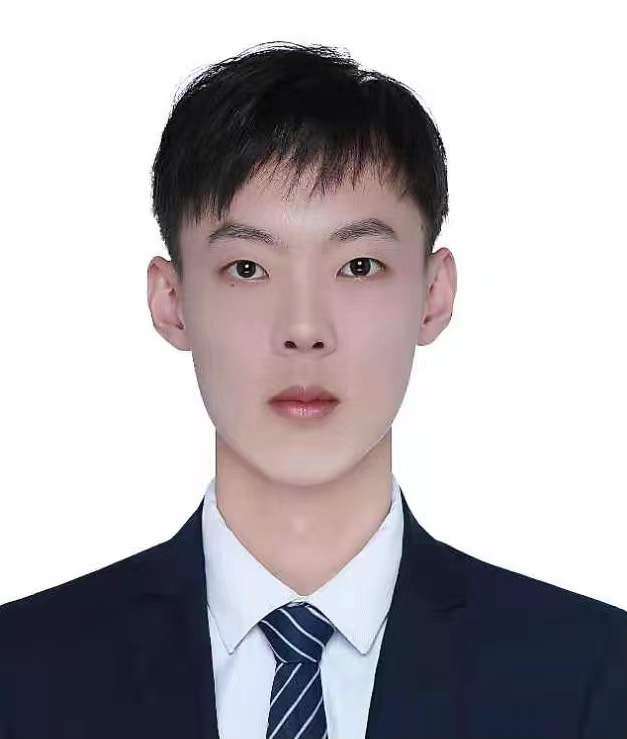}}] {Hao Geng} (Senior Member, IEEE) Currently, I am a Ph.D. student in the School of Computer Science and Engineering at Beihang University, and a member of KTL Lab, led by Professor De-Qing Wang and Fu-Zhen Zhuang. I received my B.Sc. degree from the School of Computer Science and Technology at Soochow University in June 2021. In September 2021, I was admitted to the direct Ph.D. program (BS-to-PhD) at Beihang University, under the supervision of Professor De-Qing Wang. 
\end{IEEEbiography}

\begin{IEEEbiography} [{\includegraphics[width=1in,height=1.25in,clip,keepaspectratio]{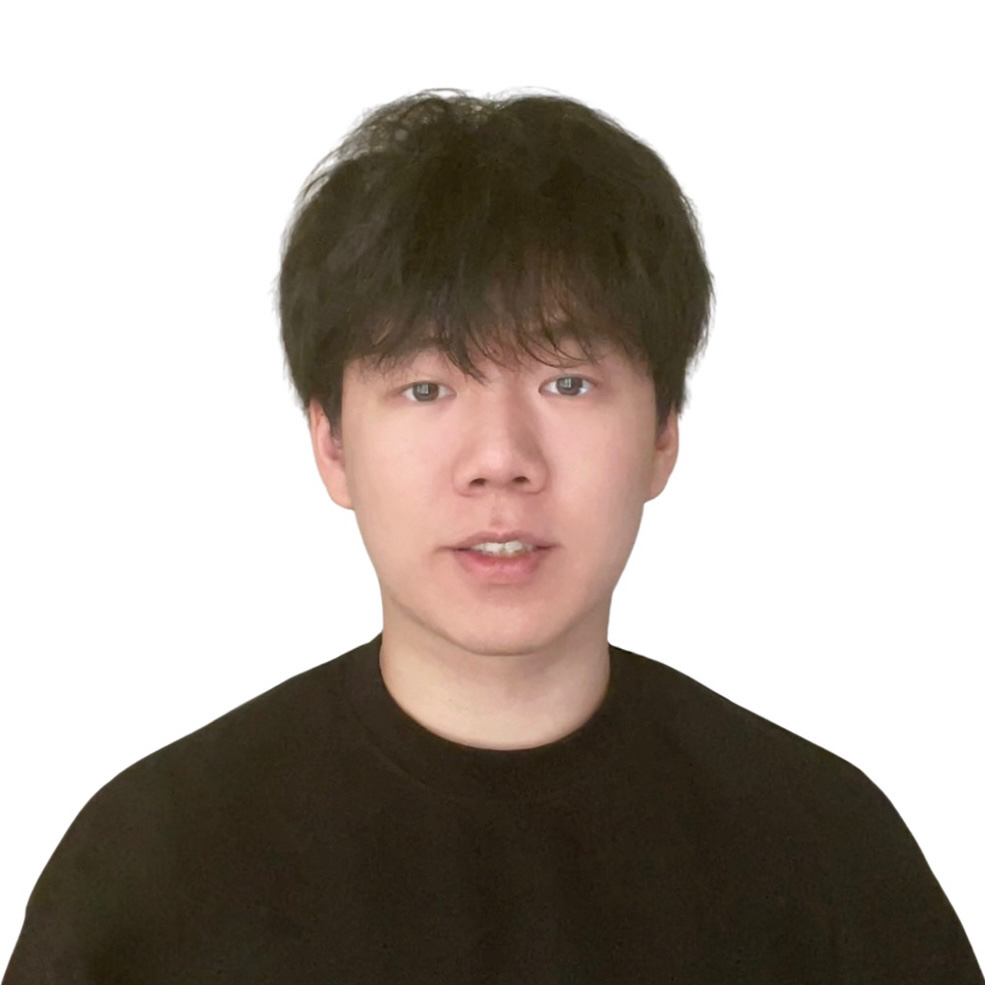}}]{Chenghao Zhang} received his Master’s degree from the University of Pennsylvania, under the supervision of Prof. Dan Roth. He is currently an Applied Scientist at Kuaishou, focusing on reinforcement learning and recommender systems.
\end{IEEEbiography}

\begin{IEEEbiography} [{\includegraphics[width=1in,height=1.25in,clip,keepaspectratio]{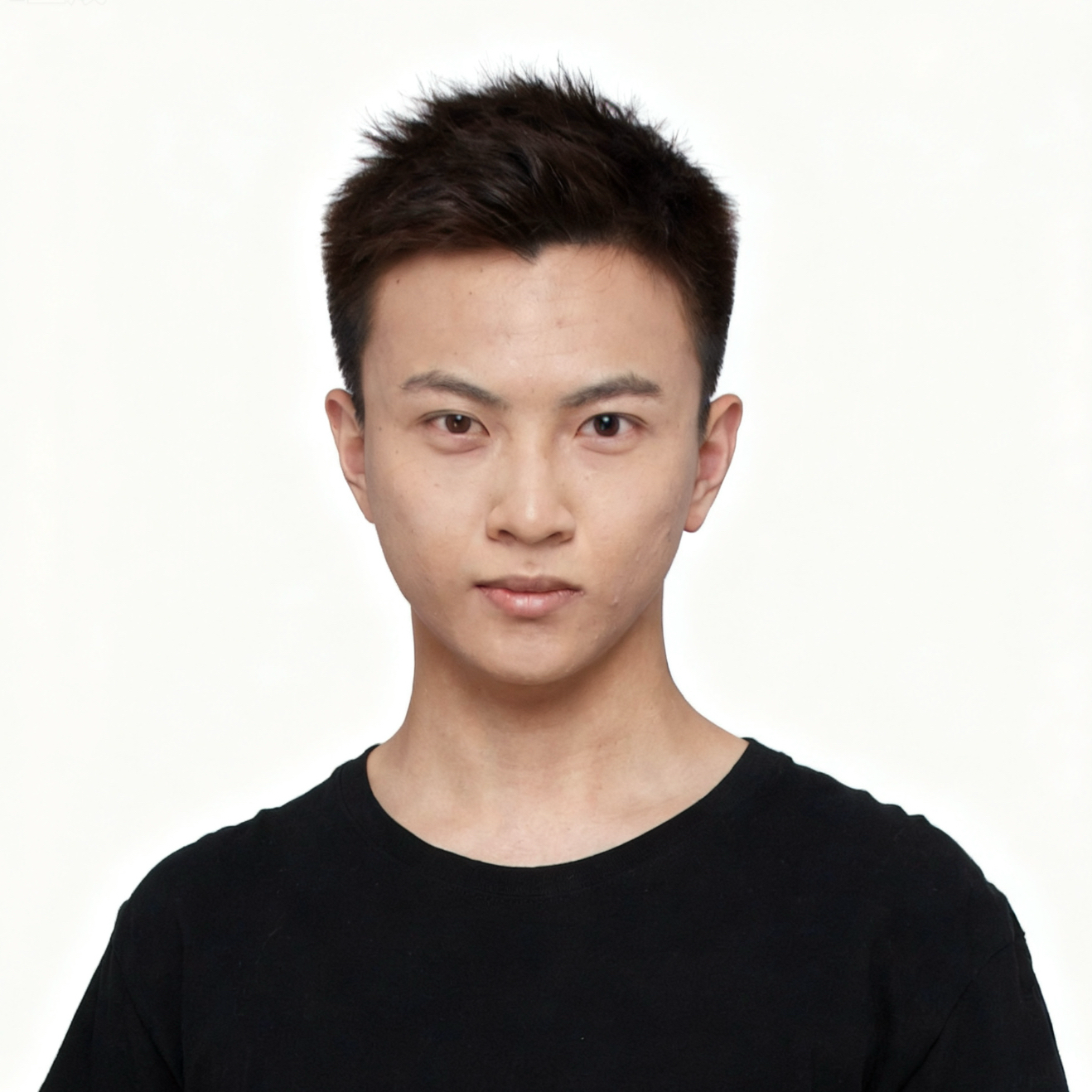}}]{Yanze Zhang} graduated from Tsinghua University in June 2022 with a Master’s degree in Electrical Engineering. I’m currently interested in machine learning/deep learning and its applications to various real-world problems, such as Search, Ads, and Recommender Systems.
\end{IEEEbiography}

\begin{IEEEbiography} [{\includegraphics[width=1in,height=1.25in,clip,keepaspectratio]{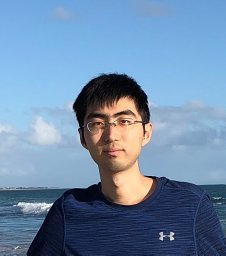}}]{Shanqi Liu} received the Ph.D. degree from Zhejiang University in 2024. He is currently a researcher at MiniMax, working on large language models. His research interests include reinforcement learning, multi-agent learning, large language models, and intelligent agents.
\end{IEEEbiography}

\begin{IEEEbiography} [{\includegraphics[width=1in,height=1.25in,clip,keepaspectratio]{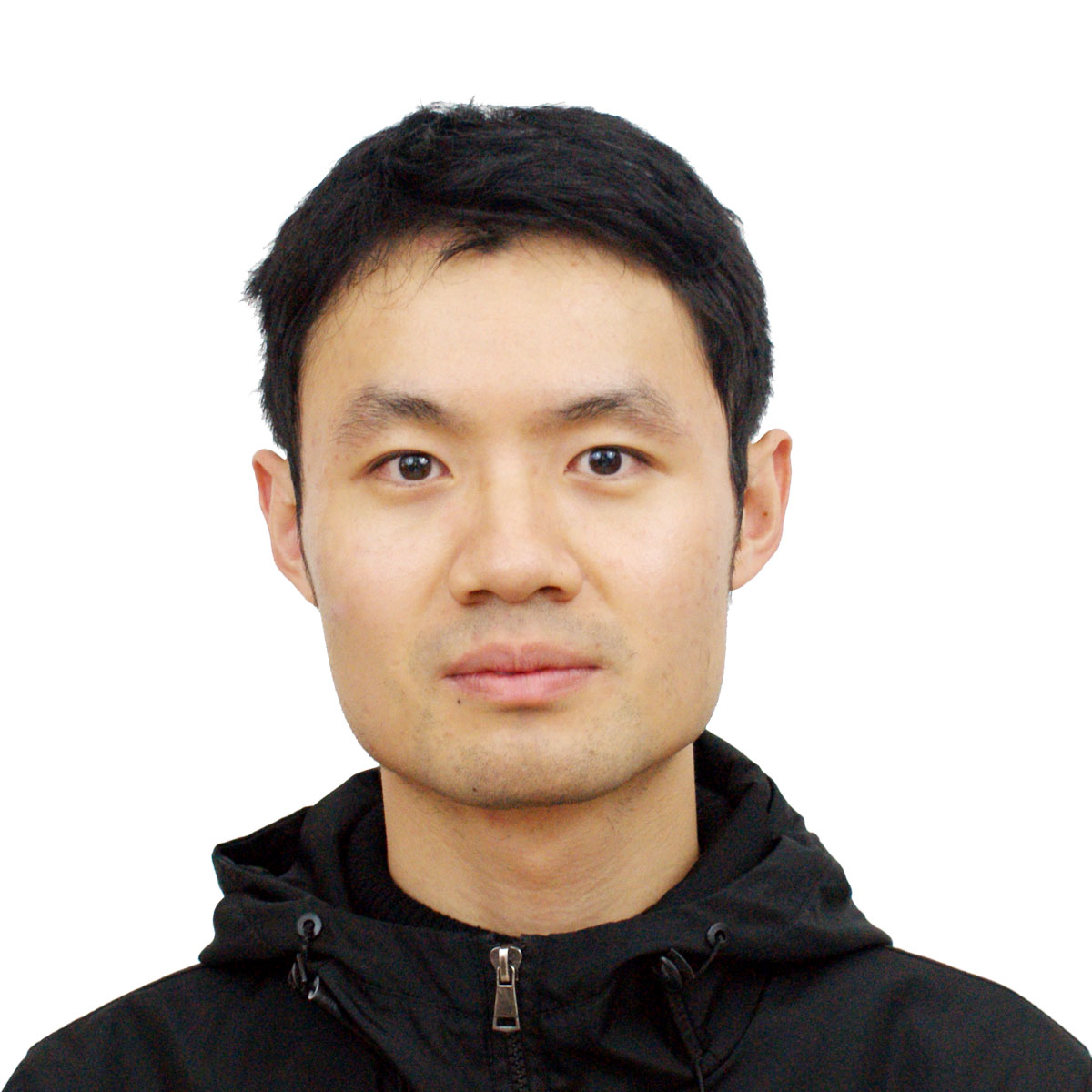}}]{Chao Feng} received his Bachelor Degree and Ph.D. degreefrom University of Science and Tenology of China in 2017 and 2023, respectively. His main research interests include Data Mining, Recommender System. He has published over 20 papers in conferences and journals including NeurIPS, KDD, TOIS, SIGIR, WWW, AAAI, IJCAI etc.
\end{IEEEbiography}

\begin{IEEEbiography} [{\includegraphics[width=1in,height=1.25in,clip,keepaspectratio]{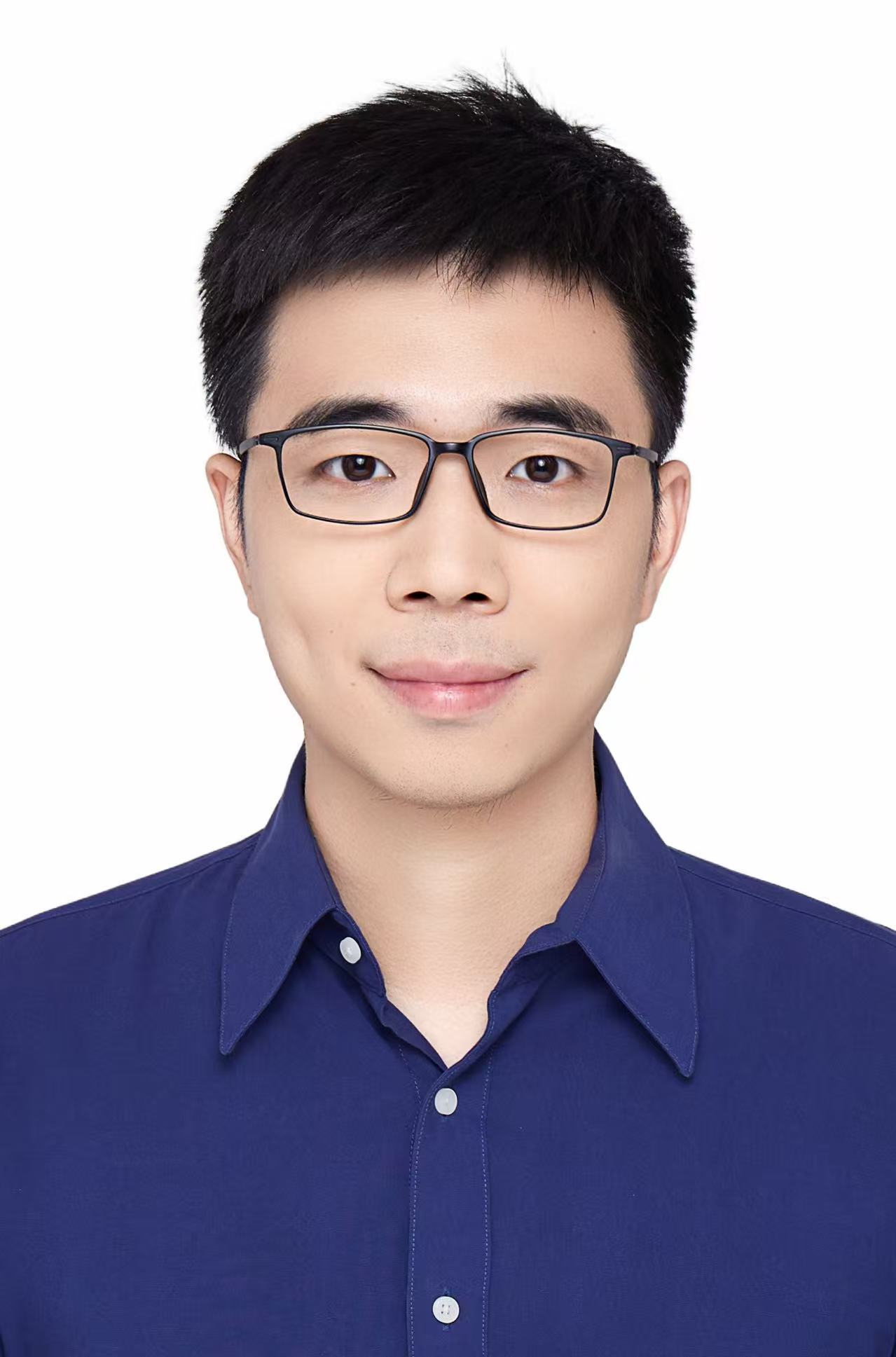}}]{Xiang Li} graduated from Tsinghua University. He is currently an Applied Scientist at Kuaishou, focusing on reinforcement learning and recommender systems.
\end{IEEEbiography}

\begin{IEEEbiography} [{\includegraphics[width=1in,height=1.25in,clip,keepaspectratio]{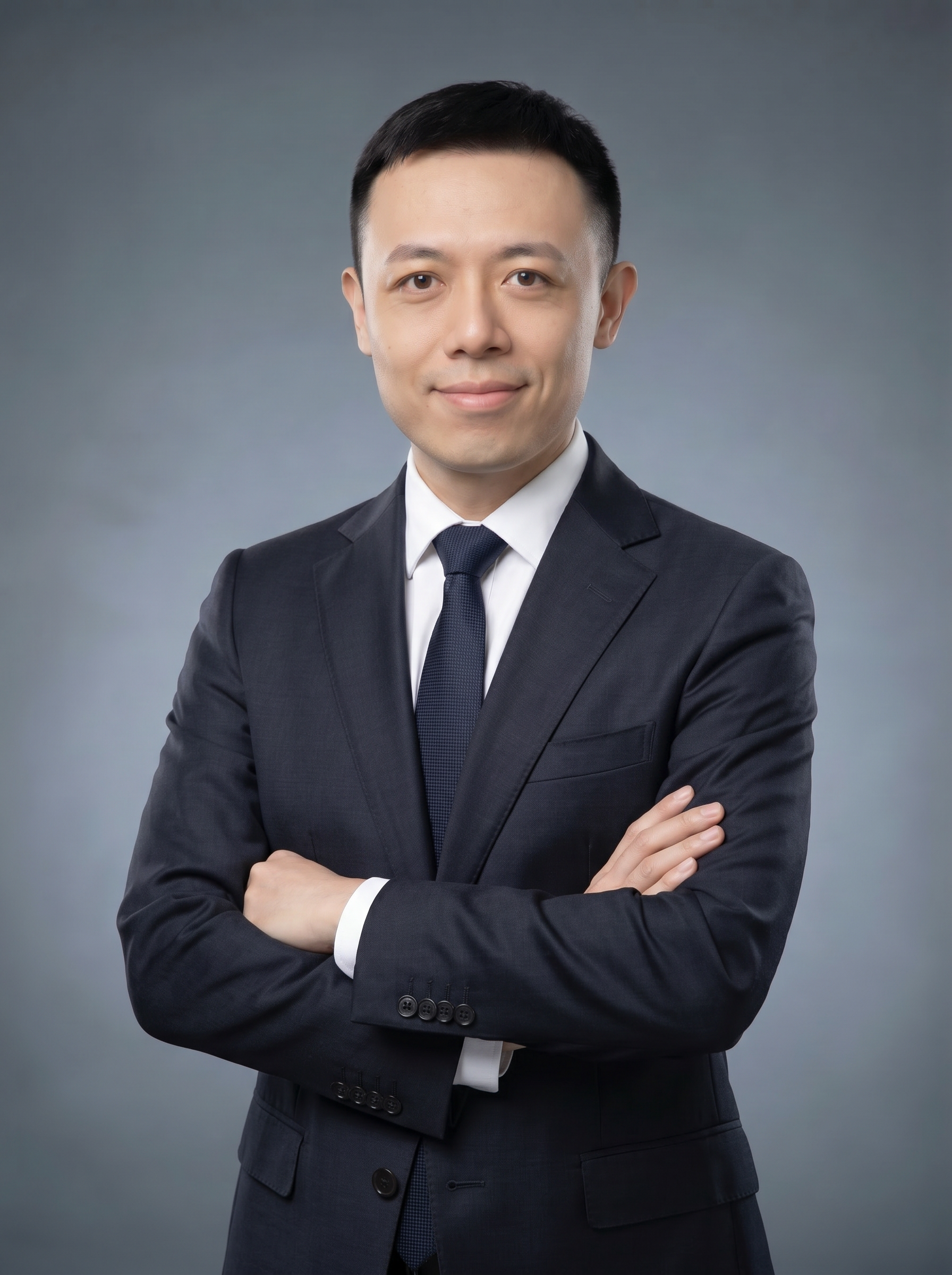}}]{Lantao Hu} graduated from Tsinghua University. He is the Kwai Technology Vice President, focusing on recommendation systems. His research interests include reinforcement learning, recommendation systems, and large language models.
\end{IEEEbiography}

\begin{IEEEbiography} [{\includegraphics[width=1in,height=1.25in,clip,keepaspectratio]{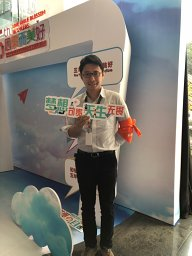}}]{Han Li} is the Kwai Technology Vice President. He holds a bachelor's and a master's degree in automation from Tsinghua University and worked at Alibaba Group in his early career. In 2019, Li Han joined Kuaishou.
\end{IEEEbiography}

\begin{IEEEbiography} [{\includegraphics[width=1in,height=1.25in,clip,keepaspectratio]{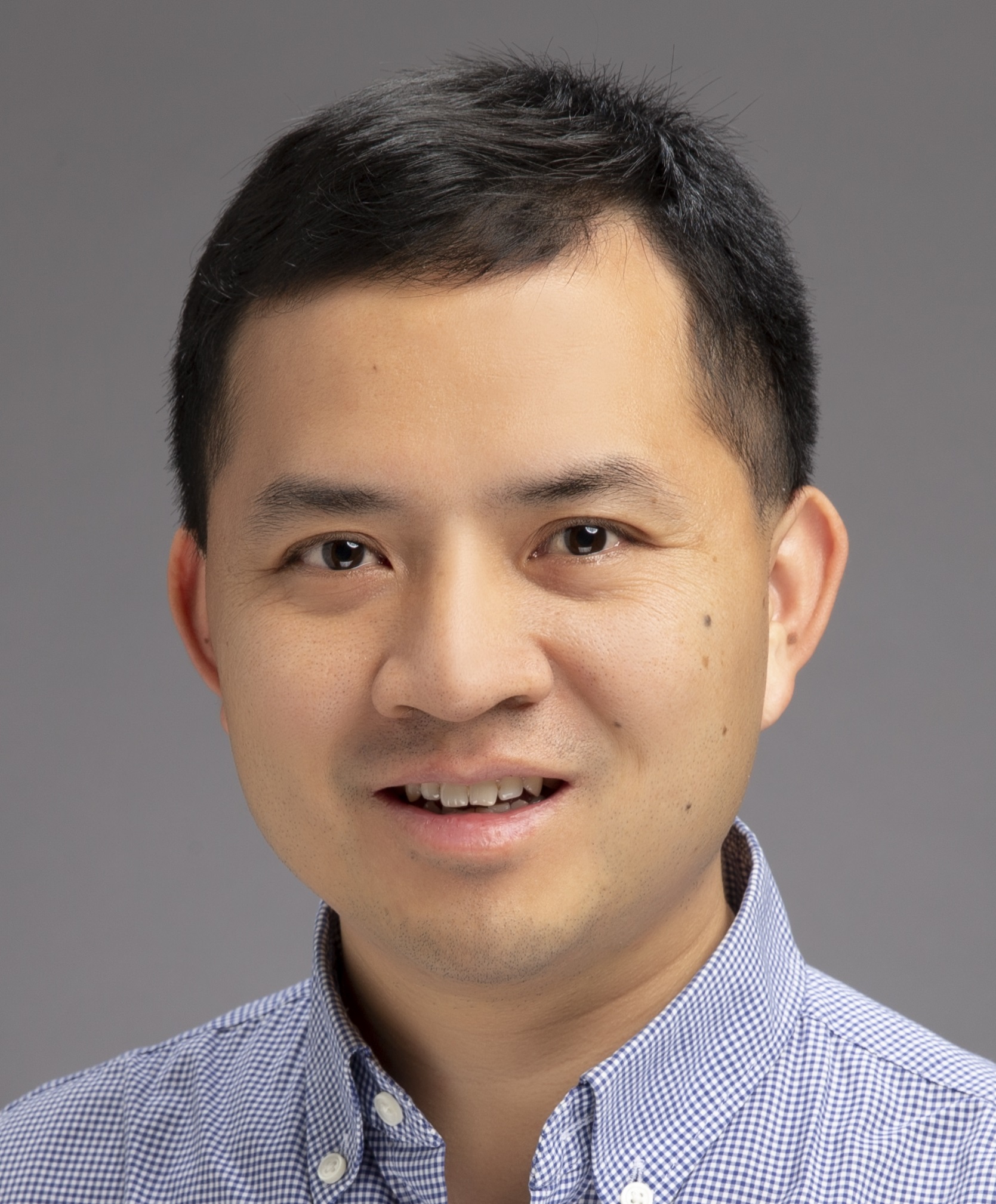}}]{Fuzhen Zhuang} is a professor with the Institute of Artificial Intelligence, Beihang University. His research interests include transfer learning, machine learning, data mining, multi-task learning and recommendation systems. He has published more than 100 papers in the prestigious refereed journals and conference proceedings, such as Nature Communications, IEEE Transactions on Knowledge and Data Engineering, Proceedings of the IEEE, IEEE Transactions on Neural Networks and Learning Systems, TIST, KDD, WWW, SIGIR, NeurIPS, AAAI.
\end{IEEEbiography}

\end{document}